# PAIRWISE VELOCITIES OF GALAXIES IN THE CFA AND SSRS2 REDSHIFT SURVEYS

Ronald O. Marzke[1,2], Margaret J. Geller[2], L.N. da Costa[3,4] and John P. Huchra[2]

[1]Dominion Astrophysical Observatory
Herzberg Institute of Astrophysics
National Research Council of Canada
5071 W. Saanich Rd.,Victoria,BC V8X 4M6
E-mail: marzke@dao.nrc.ca

[2]Harvard-Smithsonian Center for Astrophysics
60 Garden St., Cambridge, MA 02138
E-mail: mjg,huchra@cfa.harvard.edu

[3]Departamento de Astronomia CNPq/Observatorio Nacional
Rua General Jose Cristino, 77, Rio de Janeiro, Brazil 20.921

[4]Institut d'Astrophysique
98 bis Boulevard Arago Paris, France F75014
E-mail: ldacosta@iap.fr

## ABSTRACT

We combine the CfA Redshift Survey (CfA2) and the Southern Sky Redshift Survey (SSRS2) to estimate the pairwise velocity dispersion of galaxies $\sigma_{12}$ on a scale of $\sim 1h^{-1}\,\mathrm{Mpc}$. Both surveys are complete to an apparent magnitude limit $B(0) = 15.5$. Our sample includes 12,812 galaxies distributed in a volume $1.8 \times 10^6\, h^{-3}\,\mathrm{Mpc}^3$. We conclude:

1) The pairwise velocity dispersion of galaxies in the combined CfA2+SSRS2 redshift survey is $\sigma_{12} = 540\,\mathrm{km\,s^{-1}} \pm 180\,\mathrm{km\,s^{-1}}$. Both the estimate and the variance of $\sigma_{12}$ significantly exceed the canonical values $\sigma_{12} = 340 \pm 40$ measured by Davis & Peebles (1983) using CfA1.

2) We derive the uncertainty in $\sigma_{12}$ from the variation among subsamples with volumes on the order of $7 \times 10^5\ h^{-3}\ \mathrm{Mpc}^3$. This variation is nearly an order of magnitude larger than the formal error, 36 $\mathrm{km\,s^{-1}}$, derived using least-squares fits to the CfA2+SSRS2 correlation function. This variation among samples is consistent with the conclusions of Mo *et al.* (1993) for a number of smaller surveys and with the analysis of CfA1 by Zurek *et al.* (1994).

3) When we remove Abell clusters with $R \geq 1$ from our sample, the pairwise velocity dispersion of the remaining galaxies drops to $295\,\mathrm{km\,s^{-1}} \pm 99\,\mathrm{km\,s^{-1}}$.

4) The dominant source of variance in $\sigma_{12}$ is the shot noise contributed by dense virialized systems. Because $\sigma_{12}$ is pair-weighted, the statistic is sensitive to the few richest systems in the volume. This sensitivity has two consequences. First, $\sigma_{12}$ is biased low in small volumes, where the number of clusters is small. Second, we can estimate the variance in $\sigma_{12}$ as a function of survey volume from the distribution of cluster and group velocity dispersions $n(\sigma)$. For either a COBE-normalized CDM universe or for the observed distribution of Abell cluster velocity dispersions, the volume required for $\sigma_{12}$ to converge ($\delta\sigma_{12}/\sigma_{12} < 0.1$) is $\sim 5 \times 10^6 h^{-3}\ \mathrm{Mpc}^3$, larger than the volume of CfA2+SSRS2.

5) The distribution of *pairwise* velocities is consistent with an isotropic exponential with velocity dispersion independent of scale. The inferred *single*-galaxy velocity distribution function is incompatible with an isotropic exponential. Thus the observed kinematics of galaxies differ from those measured in the hydrodynamic simulations of Cen & Ostriker (1993). On the other hand, our observations appear to be consistent with the velocity distribution function measured by Zurek *et al.* (1994) using collisionless simulations with much higher resolution than Cen & Ostriker (1993). The large dynamic range in the simulations by Zurek *et al.* (1994) affords a more accurate treatment of galaxy interactions, which dramatically alter the dynamical evolution the galaxy distribution (Couchman & Carlberg 1992,Zurek *et al.* 1994). The agreement between the observed velocity distribution and the one predicted by these high-resolution simulations may be another clue that mergers play an important role in the evolution of galaxies and large-scale structure.

## 1 INTRODUCTION

Galaxy motions offer a glimpse of dark matter in the universe. Although individual peculiar velocities are notoriously difficult to measure directly, relative velocities of galaxy pairs can be measured from redshift surveys alone. Redshifts combine the cosmological component $H_0 r$ with the line-of-sight peculiar velocity. Because the distribution of galaxies on the plane of the sky is unaffected by peculiar motions, correlation functions (or, equivalently, power spectra) which are isotropic in real space are anisotropic in redshift space (*e.g.*, Sargent & Turner 1977, Peebles 1979, Peebles 1980, Kaiser 1987, Hamilton 1992). The degree of anisotropy in redshift space measures the low-order moments of the peculiar velocity distribution.

These moments play an important role in the evolution of large-scale structure. The first moment of the pairwise velocity distribution,$\bar{v}_{12}$, governs the growth of the spatial correlation function through the conservation of particle pairs (Davis & Peebles 1977, Peebles 1980, Fisher *et al.* 1994b). If galaxies trace mass and density fluctuations are in the linear regime, then $\bar{v}_{12}$ is proportional to $\Omega^{0.6}$. Fisher *et al.* (1994b) demonstrate a technique to derive $\bar{v}_{12}$ from redshift surveys using N-body simulations, and then measure $\bar{v}_{12}$ from the IRAS 1.2 Jy survey.

The second moment of the pairwise velocity distribution, $\sigma_{12}$, measures the kinetic energy of random motions in the galaxy distribution; in equilibrium, this dispersion exactly balances the gravitational potential. (Geller & Peebles 1973, Peebles 1976, Davis & Peebles 1977). Peebles (1976) measured the anisotropy of the redshift-space correlation function and derived $\sigma_{12}$ from the Reference Catalog of Bright Galaxies (deVaucouleurs & deVaucouleurs 1964). Davis & Peebles (1983, hereafter DP83) extended this analysis to a much larger redshift survey, CfA1 (Huchra *et al.* 1983). DP83 found an exponential pairwise velocity distribution with $\sigma_{12} = 340 \pm 40\,\mathrm{km\,s^{-1}}$. This observation has since become one of the strongest constraints on models for the evolution of large-scale structure, and has remained the standard against which N-body simulations are judged.

The favored paradigm for the origin of structure follows the gravitational collapse of dense regions from initially small perturbations to the large, non-linear structures we observe today. Among the many possible constituents of the initial density field, cold dark matter (CDM) stands as the theoretical standard of comparison. In its standard form, CDM presumes a flat universe; the only free parameter in the model is the normalization of the density power spectrum, usually cast as the rms density fluctuation in 8 $h^{-1}$ Mpc spheres, $\sigma_8$. Both the quadrupole anisotropy in the CMB (Wright *et al.* 1992) and the distribution of optically selected galaxies (Vogeley *et al.* 1994, Loveday *et al.* 1993) suggest that $\sigma_8 \approx 1$. Using this normalization, N-body simulations of a flat CDM universe yield $\sigma_{12} \approx 1000\,\mathrm{km\,s^{-1}}$, much larger than the values obtained by DP83. Open ($\Omega h = 0.2$) models predict $\sigma_{12} \sim 500\,\mathrm{km\,s^{-1}}$, thus reducing but not eliminating the discrepancy (Davis *et al.* 1985). Under the simple assumption that galaxy velocities trace the mass distribution, the apparent quiescence of the observed velocity field strongly constrains CDM models.

A fair comparison between theory and observation requires an accurate prescription for galaxy formation. The possibility remains that the galaxies we observe paint a biased picture of the matter distribution as a whole. If, for example, galaxies form only in dense regions, then galaxies naturally cluster more than the underlying density field (Kaiser 1984, Bardeen *et al.* 1986). This density bias lowers the normalization of the mass power spectrum corresponding to the observed galaxy power spectrum, and thus lowers the expected peculiar velocities. In simulations of a biased CDM universe, Davis *et al.* (1985) calculate $\sigma_{12}(0.5h^{-1}\,\mathrm{Mpc}) \sim 700\,\mathrm{km\,s^{-1}}$.

The COBE normalization is more restrictive. Anisotropies in the microwave background trace fluctuations in the potential itself; thus the COBE measurement removes the freedom in $\sigma_8$ allowed by density bias alone. However, Carlberg (1991) suggests that galaxy velocities may be biased as well. Carlberg (1991,1994), Carlberg & Couchman (1989), Evrard *et al.* (1992), Katz *et al.* (1992), and Cen & Ostriker (1993) describe a number of processes which rob energy from galaxy tracers either through purely gravitational interactions or through dissipation in the baryonic component. On the whole, these biases reduce the simulated $\sigma_{12}$ by up to a factor of $\sim 2$ on scales smaller than a megaparsec, not quite enough to bring the models into agreement with DP83.

The range of theoretical possibilities prompts a re-examination of the observed $\sigma_{12}$. Table 1 summarizes earlier work. The variations in the observed $\sigma_{12}$ are large. Although the largest surveys to date, CfA1 (Huchra *et al.* 1983) and IRAS12 (Fisher *et al.* 1994b), are consistent with each other to within the quoted errors, Mo *et al.* (1993) find that estimates of $\sigma_{12}$ from surveys of this size are extremely sensitive to the treatment of the few rich clusters they contain. Zurek *et al.* (1994) reach similar conclusions in their comparison between CfA1 and a series of large, high-resolution simulations. Zurek *et al.* (1994) find substantial fluctuations among simulated volumes on the scale of CfA1. Furthermore, both Mo *et al.* (1993) and Zurek *et al.* (1994) claim that CfA1 is, in fact, consistent with simulations if the redshifts are not corrected for infall into the Virgo cluster.

Here, we use a sample of 12,812 galaxies drawn from the CfA Redshift Survey (Huchra *et al.* 1995, CfA2 hereafter) and the Second Southern Sky Redshift Survey (da Costa *et al.* 1994, SSRS2 hereafter) to measure $\sigma_{12}$ and to examine the variance of $\sigma_{12}$ in volumes of $\sim 10^6 h^{-3}\mathrm{Mpc}^3$. In §2, we describe the data. In §3, we review the method we use to extract $\sigma_{12}$ from the anisotropy of the correlation function. We test the method in §4 using N-body and Monte-Carlo simulations of the galaxy distribution. Section 5 summarizes the measurements of $\sigma_{12}$ using the distribution functions discussed in §3. In §6, we discuss the implications for cosmological models and examine the convergence of $\sigma_{12}$ with survey volume by relating $\sigma_{12}$ to the distribution of group and cluster velocity dispersions. We conclude in §7.

## 2 DATA

The extension of the CfA Redshift Survey (deLapparent *et al.* 1986, Geller & Huchra 1989, Huchra *et al.* 1994, hereafter CfA2) to $B(0) \leq 15.5$ now includes approximately 11,000 galaxies. We analyze two subsamples, one in the north Galactic cap (CfA2N) and one in the south Galactic cap (CfA2S). CfA2N contains 6,480 galaxies over the region $8^h < \alpha < 17^h$, $8.5° < \delta < 44.5°$. CfA2S includes 4095 galaxies over the region $20^h < \alpha < 4^h$ and $-2.5° < \delta < 42°$. Because the survey was incomplete when we began this analysis, we do not include the northernmost slice of CfA2S included in Huchra *et al.* 1994. Because CfA2S suffers heavy extinction along its boundaries, we exclude the following regions from our analysis (cf. Vogeley *et al.* 1994): $21^h \leq \alpha \leq 4^h; 21^h \leq \alpha \leq 2^h$ and $b^{II} \geq -25°; 2^h \leq \alpha \leq 3^h$ and $b^{II} \geq -45°$, where $b^{II}$ is the galactic latitude. The remaining CfA2S sample covers 0.77 steradians and includes 2741 galaxies.

We draw CfA2 galaxies from the Zwicky catalog. Although better photometry exists for a small fraction of these galaxies, we use the original Zwicky magnitudes for all galaxies. A small fraction of the galaxies in our sample are multiple systems for which Zwicky *et al.* (1961-1968) estimated only a combined magnitude. For these cases, we estimate the relative contributions of each component by eye, and eliminate galaxies which fall below the magnitude limit.

We correct heliocentric redshifts for the solar motion with respect to the centroid of the Local Group: $\Delta v = 300 \sin(l) \cos(b)\,\mathrm{km\,s^{-1}}$. We do not correct for infall into the Virgo cluster. We exclude galaxies with redshifts smaller than 500 $\mathrm{km\,s^{-1}}$ and larger than 12,000 $\mathrm{km\,s^{-1}}$. Beyond 12,000 $\mathrm{km\,s^{-1}}$, fewer than 1% of the galaxies visible at 500 $\mathrm{km\,s^{-1}}$ are brighter than the magnitude limit given the luminosity function for CfA2N (Marzke *et al.* 1994). The mean uncertainty in the measured redshifts is 35 $\mathrm{km\,s^{-1}}$.

The Second Southern Sky Redshift Survey (da Costa *et al.* 1994, hereafter SSRS2) includes 3591 galaxies over the region $-40° < \delta < -2.5°$, $b^{II} < -40°$. Alonso *et al.* (1993a,1993b) selected the sample from a combination of two machine-generated two-dimensional catalogs: the Automatic Plate Measuring (APM) Galaxy Survey (Maddox *et al.* 1990) and the non-stellar object list of the STScI Guide Star Catalog (GSC)(Lasker *et al.* 1990, and references therein). Alonso *et al.* (1993a) calibrated the magnitude scale using overlapping galaxies in the Surface-Photometry Catalogue of the ESO-Uppsala Galaxies (Lauberts & Valentijn 1989). The ESO catalog also provided the bright galaxies missed by the automatic identification algorithms; these galaxies comprise 10% of the sample. For the SSRS2, the mean uncertainty in the redshifts is $\sim 50\,\mathrm{km\,s^{-1}}$.

These samples demonstrate rich structure in the nearby universe (Geller & Huchra 1989, da Costa *et al.* 1994). CfA2N contains the Great Wall, which includes the Coma cluster at $\sim 7,000\,\mathrm{km\,s^{-1}}$. This sample also includes the Virgo cluster nearby and several large voids in between. The most striking feature of CfA2S is the dense wall connecting the Perseus and Pisces clusters. A similar wall appears in the SSRS2 running diagonally across the survey.

Table 2 summarizes the basic features of these samples. The combined CfA2+SSRS2 sample includes 12,812 galaxies in an effective volume of $1.8 \times 10^6\,h^{-3}\,\mathrm{Mpc}^3$.

## 3 METHOD

In this section, we outline the method used to extract the pairwise velocity dispersion, $\sigma_{12}$, from redshift survey data. The technique was pioneered by Geller & Peebles (1973) and honed by Peebles (1976,1979), Davis *et al.* (1978) and Davis & Peebles (1983). Briefly, the method requires three steps: 1) establishing the variation of the redshift-space correlation function (CF, hereafter) with angle to the line of sight, 2) measuring the spatial CF from the projected separations and mean redshifts of pairs and 3) modeling the redshift-space CF as a convolution of the spatial CF with the distribution of relative peculiar velocities for pairs of galaxies.

### *3.1 Definition and Computation of $\xi(r_p, \pi)$*

To isolate distortions in the CF caused by peculiar velocities, we divide the pair separation in redshift space,$\mathbf{s}$, into two components, one along and the other perpendicular to the line of sight to the pair (Davis, Geller & Huchra 1978). The line of sight, $\mathbf{l}$, bisects the pair: $\mathbf{l} = \frac{1}{2}(\mathbf{s}_1 + \mathbf{s}_2)$ where $\mathbf{s}_i = cz_i\hat{\mathbf{r}}_i/H_0$, $c$ is the speed of light and the redshift $z \ll 1$. Assuming that the angular separation of the pair is small, the components of $\mathbf{s} = \mathbf{s}_1 - \mathbf{s}_2$ are

$$\pi = \frac{\mathbf{s}\cdot\mathbf{l}}{|\mathbf{l}|}$$
$$r_p = \sqrt{|\mathbf{s}|^2 - \pi^2} \tag{1}$$

Throughout this paper, we express the line-of-sight component $\pi$ in units of distance. For pairs with small angular separations, $\pi$ measures a combination of the Hubble separation, $\mathbf{r}_{12}$, and the component of the pairwise velocity ,$v_{12}$, along $\mathbf{r}_{12}$: $\pi \approx |\mathbf{r}_{12}| + (\mathbf{v}_{12}\cdot\mathbf{r}_{12}/|\mathbf{r}_{12}|)$. Following DP83, we restrict our analysis to pairs separated by angles smaller than $\theta_{max} = 50°$. This choice of $\theta_{max}$ balances the need for densely sampled data against the fact that widely separated pairs overestimate $v_{12}$ in the plane-parallel approximation (1).

The two-dimensional CF ,$\xi(r_p, \pi)$, is defined by the probability that any two galaxies in the survey are separated by $r_p$ and $\pi$ (Peebles 1980):

$$\delta P = n[1 + \xi(r_p, \pi)]\delta\pi\delta A \tag{2}$$

where $\delta A = 2\pi r_p \delta r_p$. To measure $\xi(r_p, \pi)$, we compute the distribution of pair separations in the data and in a Poisson realization of the data with the same radial and angular selection criteria (Peebles 1980):

$$1 + \xi(r_p, \pi) = \frac{DD(r_p, \pi)}{DR(r_p, \pi)}\frac{n_R}{n_D} \tag{3}$$

where DD and DR are the number of data/data and data/random pairs, respectively, with separations $r_p$ and $\pi$. We compute the mean densities of the real and random samples,$n_D$ and $n_R$, using the minimum-variance estimator derived by Davis & Huchra (1982).

In order to use the estimator in Equation 3, we need to choose appropriate bins in $r_p$ and $\pi$. Linear binning in both dimensions provides a clear picture of various contributions to the redshift-space anisotropy (cf. Fisher *et al.* 1994b). On the other hand, our prior knowledge of the form of the CF in $r_p$ and $\pi$ suggest a mixture of logarithmic bins in $r_p$ and linear bins in $\pi$ would be more appropriate (DP83). For our analysis, we use logarithmic bins in $r_p$ and linear bins in $\pi$. In order to facilitate comparisons with other surveys, we display the CFs using linear bins in both dimensions; again, we use this binning for display purposes only. We compute the correlation function on a 30×30 grid with 1 Mpc bins in $\pi$ and 0.1dex bins in $r_p$. The 30 bins in $r_p$ cover the range 0.1 Mpc to 100 Mpc. To fit models to $\xi(r_p, \pi)$, we use coarser binning in $r_p$ to increase the signal-to-noise: each bin represents a factor of 2 in $r_p$ starting at $r_p = 0.1$ Mpc.

To construct the random samples, we use the luminosity functions derived by Marzke *et al.* (1994) for the CfA survey and by da Costa *et al.* (1994) for the SSRS2. Because we lack morphologies for most galaxies in the sample, we use the general luminosity function averaged over all types. The luminosity functions for CfA2N, CfA2S and SSRS2 differ significantly. In order to remain self-consistent, we construct the random sample for CfA2+SSRS2 using the individual luminosity functions rather than using a mean averaged over all samples. Although these variations in the luminosity function complicate the interpretation of large-scale features in $\xi(r_p, \pi)$, they do not affect our analysis of the small-scale velocity field.

To compute the pair sums DD and DR, galaxies may be counted in any number of ways as long as the weight assigned to each galaxy does not depend explicitly on the local density (Hamilton 1993). In a "fair" sample, almost by definition, any weighting yields an unbiased estimate. The simplest approach is to count each galaxy equally. Because the density of galaxies in a magnitude-limited sample drops rapidly with distance, this weighting favors nearby galaxies. Weighting by the integral of the luminosity function to account for the magnitude limit gives equal weight to each volume of space (Davis & Huchra 1982). In the minimum-variance (MV, hereafter) estimate of $\xi(r_p, \pi)$ for a fair sample of galaxies (Efstathiou *et al.* 1988), each pair carries a weight $w_1 w_2$ where

$$w_i = \frac{1}{1 + 4\pi n \phi(s_i) J_3(s_{ij})} \tag{4}$$

Here, $s_i$ is the redshift of galaxy $i$, $s_{ij}$ is the pair separation in redshift space, $\phi$ is the selection function and $J_3$ is the volume integral of the spatial CF (Peebles 1980). MV weighting balances the advantage of equal-volume weighting against the shot-noise caused by the few close pairs at large redshift (Efstathiou *et al.* 1988, Loveday *et al.* 1992, Park *et al.* 1994,Fisher *et al.* 1992).

The scale of the CfA2+SSRS2 redshift survey is no larger than the scale of the largest apparent structures. In these samples, different weighting schemes yield systematically different estimates of the CF, depending on the variation of clustering with distance and with galaxy luminosity (deLapparent *et al.* 1988, Park *et al.* 1994). We show in §5 that variations in $\xi(r_p,\pi)$ derived with different weighting schemes are not negligible. Because MV weighting ensures the most efficient use of the data, we present our final numerical results using this weighting exclusively.

Correlation statistics of samples containing structure on the scale of the survey are particularly sensitive to uncertainty in the mean density of the universe (deLapparent *et al.* 1988, Kaiser 1987, Hamilton 1993). Because the mean is most often defined internally as the number of sample galaxies divided by the sample volume, the volume integral of the CF vanishes artificially on the scale of the survey. On large scales, where the CF is small, errors in the CF are proportional to the error in the mean density, $\delta$ (deLapparent *et al.* 1988). Hamilton (1993) suggests an alternative to the standard estimator (Equation 3) which is less affected by uncertainty in the mean density:

$$1+\xi_H(r_p,\pi) = \frac{\langle DD\rangle\langle RR\rangle}{\langle DR\rangle^2} \tag{5}$$

Because the error in $\xi_H$ is proportional to $\delta^2$, this estimator performs better than Equation 3 when $\delta \ll 1$. In the samples we analyze here, $\delta$ is not much smaller than unity (Marzke *et al.* 1994, da Costa *et al.* 1994), and $\xi_H \approx \xi$. For our purposes, the differences between the estimators in Equations 3 and 5 are negligible. Fisher *et al.* (1994b) reach similar conclusions in their analysis of the IRAS 1.2Jy survey.

### *3.2 Model for $\xi(r_p,\pi)$*

$\xi(r_p,\pi)$ may be expressed as a convolution of the spatial CF, $\xi(r)$, with the distribution of pairwise peculiar velocities (Peebles 1980):

$$1+\xi(r_p,\pi) = \int [1+\xi(r)]P(\mathbf{v}_{12}|\mathbf{r}_{12})\,d^3\mathbf{v}_{12} \tag{6}$$

If the pairwise peculiar velocity distribution varies slowly with pair separation and if there is no preferred direction in the velocity field, then $\xi(r_p,\pi)$ may be expressed in terms of the distribution of line-of-sight velocities alone:

$$1+\xi(r_p,\pi) = \int [1+\xi(r)]P(v_{los}|r_{12})\,dv_{los} \tag{7}$$

If we similarly decompose the spatial separation $\mathbf{r}_{12}$ into components $(r_p, y)$ perpendicular to and along the line of sight, then $r_{12}^2 = r_p^2 + y^2$, $v_{los} = \pi - y$ and

$$1+\xi(r_p,\pi) = \int [1+\xi(\sqrt{r_p^2+y^2})]P((\pi-y)|r_{12})\,dy \tag{8}$$

(Peebles 1980).

This expression is particularly simple when both members of the pair share the same line of sight, i.e. $r_p = 0$. In this case,

$$1+\xi(r_p=0,\pi) = \int [1+\xi(y)]f((\pi-y)|y)dy \tag{9}$$

Thus if we know the spatial CF and we measure $\xi(r_p,\pi)$, we can probe the peculiar velocity distribution. Specifically, we measure the distribution function of a single component of the pairwise velocity, $P(v_{12}^{\alpha})$. We will see later that the cut in $\xi(r_p,\pi)$ at $r_p = 0$ best discriminates among models for $P(v_{12}^{\alpha})$.

### *3.3 Distortions in $\xi(r_p,\pi)$ from Peculiar Velocities*

Two limiting cases illuminate the distortions in $\xi(r_p,\pi)$ caused by peculiar velocities. On scales where the width, $\sigma_{12}$, of the pairwise velocity distribution exceeds the Hubble separation (for example in relaxed clusters and groups), pairs that are close in space are on average more widely separated in redshift space. Thus the slope of the redshift-space CF is shallower along the line of sight than perpendicular to it; the contours of $\xi(r_p,\pi)$ are extended along the line of sight. This distortion is the familiar "redshift finger", a distinctive feature of galaxy redshift surveys. The length of the fingers is a standard measure of the second moment of the velocity distribution function.

Another distortion is caused by the mean motion of galaxies toward one another. On scales where the mean infall velocity is smaller than the Hubble separation, infall compresses $\xi(r_p,\pi)$ along the line of sight (Peebles 1980). Because density fluctuations on these scales are small, linear perturbation analysis yields an expression relating the amplitude of this distortion to the mean mass density of the universe $\Omega$ (Peebles 1980, Kaiser 1987, Hamilton 1993, Fisher *et al.* 1994b). Given a model for the bias between galaxies and mass, the mean infall velocity provides a direct route to $\Omega$. Unfortunately, coherent structure on the scale of the survey masquerades as large-scale streaming (Fisher *et al.* 1994b), and the true infall velocity is nearly impossible to disentangle.

### *3.4 Spatial Correlation Function*

Our ability to explore the pairwise velocity distribution through Equation 8 rests on our knowledge of the true distribution of galaxies in space. If we measure $\xi(r_p,\pi)$ on large enough scales, we can deproject $\xi(r)$ from $\xi(r_p,\pi)$ without modeling the velocity distribution function at all (DP83). The projection of $\xi(r_p,\pi)$ onto the $r_p$ axis, $w(r_p)$, depends only on the spatial CF:

$$\begin{aligned} w(r_p) &= \int_0^{\pi_{max}} \xi(r_p,\pi)\, d\pi \\ &= \int_0^{\infty} dy\, \xi(\sqrt{r_p^2+y^2}) \qquad r_p \ll \pi_{max} \end{aligned} \tag{10}$$

In practice, we measure $\xi(r_p,\pi)$ to a scale $\pi_{max}$ where the signal-to-noise approaches unity. The projection in Equation 10 is independent of the velocity distribution only if $\pi_{max} \gg \sigma_{12}$. Given $w(r_p)$, we invert Equation 10 for $\xi(r)$ (DP83):

$$\xi(r) = -\frac{1}{\pi}\int_r^{\infty} \frac{w(r_p)\, dr_p}{\sqrt{r_p^2 - r^2}} \qquad r \ll \pi_{max} \tag{11}$$

For a power law spatial CF $\xi(r) = (r/r_0)^\gamma$, the projected CF is $w(r_p) = Ar_p^{1+\gamma}$ with

$$A = r_0^\gamma \frac{\Gamma(1/2)\Gamma[(\gamma-1)/2]}{\Gamma(\gamma/2)}. \tag{12}$$

*3.5 Velocity Distribution Function*

In linear theory, the peculiar velocity field of individual galaxies is simply related to the density field (Peebles 1980):

$$\mathbf{v}_1(\mathbf{r}) = \frac{H_0\Omega^{0.6}}{4\pi}\int \frac{(\mathbf{r}-\mathbf{r}')\delta(\mathbf{r}')}{|\mathbf{r}-\mathbf{r}'|^3}\, d^3r' \tag{13}$$

If the one-point density distribution is Gaussian, then the velocity distribution function is Gaussian and is completely specified by the power spectrum of density fluctuations (Vittorio *et al.* 1989, Scherrer 1992):

$$\sigma_1^2 = H_0^2\Omega^{1.2}\int_0^\infty 4\pi P(k)\, dk \tag{14}$$

On the small scales we observe here, non-linear evolution of the phase-space distribution erases much of the history written in the initial galaxy orbits. In this regime, the velocity distribution is difficult to predict. Our knowledge of the small-scale velocity distribution comes primarily from N-body simulations and is further guided by observations. Peebles (1976) calculated $\xi(r_p, \pi)$ using the Reference Catalogue of Bright Galaxies (deVaucouleurs and deVaucouleurs 1964) and observed that an exponential distribution of pairwise line-of-sight velocities best reproduced the data:

$$P(v_{12}^\alpha) = \exp\left(-\sqrt{2}v_{12}^\alpha/\sigma_{12}\right) \tag{15}$$

A number of later surveys confirm this conclusion (§4, Peebles 1979, DP83, Bean *et al.* 1983, Hale-Sutton *et al.* 1989, Fisher *et al.* 1994b).

Velocity distributions derived from N-body simulations vary. Efstathiou *et al.* (1988) note that the distribution of pairwise velocities in their simulations differs substantially from an exponential. The difference is particularly noticeable at small relative velocities, where $P(v_{12}^\alpha)$ is much flatter than the exponential distribution. Fisher *et al.* (1994b) confirm this behavior in the simulations they use to interpret their data.

Following both collisionless and baryonic particles in their N-body simulations, Cen and Ostriker (1993) find that the distribution of 3-d *single*-"galaxy" peculiar velocities is well represented by an isotropic exponential:

$$f(\mathbf{v}_1)d^3\mathbf{v}_1 = \exp -\beta|\mathbf{v}_1|d^3\mathbf{v_1} \tag{16}$$

where $\beta = \sqrt{2}/\sigma_1$. For this distribution, the probability of measuring a velocity with magnitude $v_1$ is proportional to $v_1^2 e^{-\beta|v_1|}$.

We would like to explore not only the moments of $P(v_{12}^\alpha)$ but also the functional form of the distribution. We can derive a fitting function for $P(v_{12}^\alpha)$ to compare with the N-body model in Cen and Ostriker (1993) as follows. If the velocity field is isotropic, the probability of measuring one component $v_1^\alpha$ of a single-galaxy velocity is

$$P(v_1^\alpha) = \int_{-\infty}^{\infty}\int_{-\infty}^{\infty} \exp\left[-\beta\left|\sqrt{(v_1^\alpha)^2 + (v_1^2)^2 + (v_1^3)^2}\right|\right] dv_1^2 dv_1^3$$
$$= (|v_1^\alpha| + \frac{1}{\beta}) \exp\left[-\beta |v_1^\alpha|\right] \qquad (17)$$

The second moment of this one-dimensional, *single*-galaxy distribution is $\sqrt{2}\sigma_1$. A similar transformation yields the distribution of one component of the *pairwise* velocity $P(v_{12}^\alpha)$, where $v_{12}^\alpha = v_1^\alpha(1) - v_1^\alpha(2)$. To simplify the notation, we define the variables $u_i = v_1^\alpha(i)$ so that $v_{12}^\alpha = u_1 - u_2$. Assuming the velocities of the galaxies are isotropic and independent,

$$P(v_{12}^\alpha) = \int_{-\infty}^{\infty} P_1(v_{12}^\alpha + u_2) P_1(u_2)\, du_2$$
$$= \int_{-\infty}^{\infty} (|v_{12}^\alpha + u_2| + \frac{1}{\beta})(|u_2| + \frac{1}{\beta}) \exp\left[-\beta(|v_{12}^\alpha + u_2| + |u_2|)\right] du_2$$
$$= \left[\frac{|v_{12}^\alpha|^3}{6} + \frac{|v_{12}^\alpha|^2}{\beta} + \frac{5}{2\beta^2}(|v_{12}^\alpha| + \frac{1}{\beta})\right] \exp\left(-\beta|v_{12}^\alpha|\right) \qquad (18)$$

The second moment of this one-dimensional pairwise distribution $P(v_{12}^\alpha)$ is $2\sigma_1$, *i.e.* twice the 3-d single-galaxy velocity dispersion.

Figure 1 shows the normalized distributions $P(v_{12}^\alpha)$ given by Equations 15 and 18. The distributions are normalized to have the same second moment; $\sigma_{12} = 500\,\mathrm{km\,s^{-1}}$ for the distribution in Equation 15 and $\sigma_1 = 250\,\mathrm{km\,s^{-1}}$ for the distribution in Equation 18. If $P(v_1)$ is exponential and we mistakenly assume that $P(v_{12}^\alpha)$ is exponential with $\sigma_{12} = 2\sigma_1$, we overestimate the number of pairs at small separations. It is important to emphasize that an exponential distribution of *single*-galaxy velocities does not imply an exponential distribution of *pairwise* velocities.

The distribution $P(v_{12}^\alpha)$ in Equation 18 is qualitatively similar to pairwise velocity distributions measured directly from N-body simulations. Efstathiou *et al.* (1988) and Fisher *et al.* (1994b) plot these distributions for a range of initial conditions and epochs, and in each case, the core of the distribution is noticeably flatter than exponential. This small-scale flattening may be partially responsible for the discrepancy observed by Fisher *et al.* (1994b) between their isotropic exponential model for $\xi(r_p, \pi)$ and the CF calculated directly from their N-body simulations. Of course, as Fisher *et al.* (1994b) and Cen and Ostriker (1993) note, pairwise velocities in the real universe and in simulations are anisotropic and correlated; thus the assumptions that go into the derivation of Equation 18 are not entirely valid. Although Equation 18 only approximates the behavior of $v_{12}$ in the simulations, it provides a useful guide to departures from the exponential distribution in Equation 15.

To summarize, we test the underlying velocity field by fitting models to $\xi(r_p, \pi)$ using both exponential *pairwise* distributions and exponential *single*-galaxy distributions. For the latter case, we use the pairwise distribution in Equation 18 for the convolution Equation 8. We test the validity of Equation 18 in §4.2. We emphasize that neither distribution is based on a physical model; Equation 15 is motivated by observations, and Equation 18 is a simple empirical model which ignores correlations and anisotropy in the velocity field. Clearly, a physical model for the velocity distribution function in the non-linear regime would be extremely useful (e.g. Kofman 1991 and references therein).

### *3.6 Scale Dependence of the Velocity Moments*

If clustering is statistically stable, the cosmic virial theorem implies that $\sigma_{12}$ scales as $r^{2-\gamma}$ where $\gamma$ is the slope of the spatial CF. Thus for $\gamma$ in the observed range ($\sim$ 1.8–1.9), $\sigma_{12}$ is nearly constant with scale. On large scales, linear theory requires that $\sigma_{12}$ decrease slowly with scale (Peebles 1980). The scaling of $\sigma_{12}$ in N-body simulations varies according to the details of the model and the algorithm. Fisher *et al.* (1994b) adjust the scaling of $\sigma_{12}$ from their N-body simulations to match the observed CF and to reproduce both the cosmic virial theorem on small scales and the linear theory prediction on large scales. Because the behavior of $\sigma_{12}$ derived by Fisher *et al.* is essentially scale-independent, and because we don't know what resemblance the simulations bear to the real universe, we choose to ignore the scale dependence of $\sigma_{12}$. We show in §4 that this choice does not hinder our ability to measure $\sigma_{12}$ on small scales even in simulations where $\sigma_{12}$ varies significantly with scale.

Our knowledge of the scaling of the first moment,$\bar{v}_{12}$, is similarly vague. Because $\sigma_{12}$ is likely to be significantly larger than $\bar{v}_{12}$ on small scales, we choose to calculate $\sigma_{12}$ without modeling the infall pattern. To investigate the effects of infall on our derivation of $\sigma_{12}$, we adopt the model used by DP83 derived from the scale-invariant solution to the truncated BBGKY heirarchy (Davis & Peebles 1977):

$$\bar{v}_{12}(r) = \frac{Fr}{1 + (r/r_0)^2} \tag{19}$$

The similarity solution implies $F = 1$ (DP83).

### *3.7 Model Fitting*

Standard $\chi^2$ analysis is inadequate for correlation analysis for two reasons. First, estimates of the CF, $\hat{\xi}$, at different separations are correlated. The correlation between points effectively reduces the number of degrees of freedom; if correlations are large, they can significantly alter the fit. Second, Fisher *et al.* (1994a) report that the distribution of $\hat{\xi}$ is significantly non-normal over portions of the $(r_p, \pi)$-plane.

In order to address the covariance between $\hat{\xi}$ at different separations, we use techniques of principal component analysis elucidated by Fisher *et al.* (1994a). We compute the covariance matrix using 100 bootstrap resamplings of the original dataset. Because the covariance matrix of $\hat{\xi}$ is real and symmetric, it can be diagonalized by an orthogonal transformation. This transformation defines a set of linear combinations of the original data points which are linearly independent and are therefore appropriate for $\chi^2$ analysis. The transformed statistic is

$$\tilde{\chi}^2 = \sum_{i=1}^{N} \frac{(\tilde{\xi}_i - \tilde{\xi}_{i,model})^2}{\tilde{\sigma}_i^2} \tag{20}$$

where $\tilde{\sigma}_i$ are the eigenvalues of the original covariance matrix and $\tilde{\xi}_i$ and $\tilde{\xi}_{i,model}$ are the transformed data and model vectors, respectively. The columns of the matrix which transforms $\xi$ to $\tilde{\xi}$ are simply the eigenvectors of the original covariance matrix $V$.

The non-normality of $\hat{\xi}$ is more problematic. Fisher *et al.* (1994a) use N-body simulations to show that this distribution is slightly but significantly skew on the scales of interest. Thus the standard $\chi^2$ distribution does not accurately represent the expected distribution of our statistic. Although this non-Gaussianity mitigates some of the statistical advantages of principal component analysis, we use $\tilde{\chi}^2$ to address the covariance between points and quote the confidence interval under the assumption of Gaussian errors. Because the error distribution is likely to be broader than a Gaussian, we interpret borderline rejections ($P(\chi^2, \nu) \sim .01$) cautiously.

## 4 TESTS OF THE METHOD

In this section, we test the accuracy of the procedure outlined in §3 using simulated data. M. Crone and A. Evrard kindly provided an N-body simulation for this analysis. Crone and Evrard (1994) follow the motion of 262,144 CDM particles in a 64 $h^{-1}$ Mpc box using a particle-particle/particle-mesh algorithm. Initial density perturbations have a power-law spectrum $\delta(\mathbf{k}) = |\mathbf{k}|^n$ where $n = -1$, and the cosmology is flat ($\Omega = 1, \Lambda = 0$). Using a random subsample of the particles, we construct mock redshift surveys by viewing the particles from corners of the box and assigning a uniform Hubble flow. From these mock surveys, we compute $\xi(r_p, \pi)$, $w(r_p)$, and $\xi(r)$ as we would for the real data. Because the simulation volume is small, we do not sample enough independent structures to draw large numbers of mock surveys; this restriction limits our ability to test for bias in the techniques. Still, the simulation serves as a convenient and physically motivated set of known phase-space coordinates. We do not intend to test these models against the data; we use them only to establish the accuracy and the limitations of our procedure.

### *4.1 $\xi(r_p, \pi)$ from the Simulation*

The velocity field of the N-body simulation is very hot; $\sigma_{12}(0.5h^{-1}\,\mathrm{Mpc}) = 1435\,\mathrm{km\,s^{-1}}$. Because we cannot measure $\xi(r_p, \pi)$ accurately beyond 30 $h^{-1}$ Mpc, we cannot test our procedure adequately with the raw velocity field. Following Davis *et al.* (1982), we cool the velocity field by dividing all particle velocities by a factor of three. This procedure is roughly (though not precisely) equivalent to lowering the mean mass density to $\Omega \sim 0.11$ (Davis *et al.* 1982). The reduced dispersion is $\sigma_{12}(0.5h^{-1}\,\mathrm{Mpc}) \sim 500\,\mathrm{km\,s^{-1}}$.

Figure 2a shows $\xi(r_p, \pi)$ calculated from a mock survey of the simulated galaxy distribution. The elongation along the line of sight caused by peculiar velocities is clearly evident. Compression of the contours at large $r_p$ hints at the infall pattern discussed in §3.2. Figure 2b shows $\xi(r_p, \pi)$ in the absence of peculiar velocities. Although the spatial distribution is nearly isotropic on scales less than the correlation length ($\sim 5h^{-1}$ Mpc), the shape of the structures themselves distort $\xi(r_p, \pi)$ on larger scales. Because we ascribe anisotropy in $\xi(r_p, \pi)$ to peculiar velocities, this intrinsic spatial anisotropy fundamentally limits our ability to extract the velocity moments (particularly $\bar{v}_{12}$) from finite redshift surveys (Fisher *et al.* 1994b, Marzke *et al.* 1994b).

### *4.2 Inversion of $w(r_p)$ for $\xi(r)$*

Figure 3 shows how the observational cutoff $\pi_{max}$ affects the deprojection of $\xi(r)$. The solid line is a pure power law. We use this power law to make an analytic model of $\xi(r_p, \pi)$ using Equation 8 and then calculate $w(r_p)$ numerically. We then invert $w(r_p)$ for $\xi(r)$ using Equation 12. We show inversions for $\pi_{max} = 30h^{-1}$ Mpc and for $\pi_{max} = 60h^{-1}$ Mpc. Deviations from a power law are significant beyond 10 $h^{-1}$ Mpc for $\pi_{max} = 30h^{-1}$ Mpc. We therefore restrict our analysis of $\xi(r)$ to $r < 10h^{-1}$ Mpc.

Figure 4 shows the fit to $w(r_p)$ measured from the simulation. Figure 5 compares $\xi(r)$ computed from the mock survey with the CF calculated directly from pair counts using the known spatial coordinates. To the extent that the simulated CF can be represented by a power law, the inversion of $w(r_p)$ reproduces the spatial CF on scales smaller than $\sim$10 Mpc.

*4.3 Recovering a Known Velocity Field*

To test our ability to recover a known velocity field, we start with a simple model. For each particle in the N-body simulation, we replace the computed velocities with random deviates drawn from the isotropic exponential distribution in Equation 16 with $\sigma_1 = 250\,\mathrm{km\,s^{-1}}$. We preserve the clustered spatial distribution; we replace only the computed velocities. Figure 6 shows the distribution of 1-d pairwise velocities measured from the simulation along with the prediction of Equation 18. Equation 18 predicts the measured pairwise distributions well.

Figure 2c shows $\xi(r_p, \pi)$ calculated from a mock redshift survey of this model. On large scales, the characteristic compression caused by infall, which is subtle but clear in figure 2a, is clearly absent in figure 2(c) where $\bar{v}_{12} = 0$. Figure 7 shows fits to $\xi(r_p, \pi)$ for the first six broad bins in $r_p$. The solid line represents the convolution in Equation 8 with $P(v_{12}^{\alpha})$ given by Equation 18. The fit is good. We recover the input $\sigma_1$ to within one standard deviation: $\sigma_1(r_p \leq 0.8h^{-1}\,\mathrm{Mpc}) = 260 \pm 14\,\mathrm{km\,s^{-1}}$.

*4.4 Recovering $\sigma_{12}$ from the N-body Simulation*

In the N-body simulation, the velocity moments vary with pair separation (cf. Efstathiou *et al.* 1988, Fisher *et al.* 1994a). Figure 8 shows $\sigma_{12}$ and $\bar{v}_{12}$ as a function of pair separation. We measure these moments directly from the particle positions and velocities computed in the simulation. Because we assume in our model that the velocity dispersion is independent of scale, we worried that the dispersion $\sigma_{12}$ we infer from $\xi(r_p, \pi)$ might be biased. The open squares in Figure 8 show $\sigma_{12}$ measured from $\xi(r_p, \pi)$ using the model (Equation 8) and ignoring infall. The measured dispersion follows the true dispersion remarkably well. With only one exception, the measured $\sigma_{12}$ falls within one standard deviation of the true dispersion.

The dotted line in Figure 8 shows the measured $\sigma_{12}$ assuming the infall model in Equation 19 with $F = 1.5$ (this model is shown as the dashed line in Figure 8). The inclusion of infall actually degrades the measurement of $\sigma_{12}$. Although the dispersion at $r_p \leq 1h^{-1}\,\mathrm{Mpc}$ is consistent with the true dispersion, the measurements at larger $r_p$ are biased high. DP83 note a similar discrepancy in their analysis of the simulations described by Efstathiou and Eastwood (1981). The source of this discrepancy is unclear, but there are several possibile culprits: the spatial CF deviates from a power law in the N-body simulation, and in detail, $\sigma_{12}$ is neither isotropic nor scale-independent. Because we recover $\sigma_{12}$ within one standard deviation without including infall, we derive $\sigma_{12}$ for the data using the no-infall model. We show results for $\sigma_{12}$ including infall to emphasize that the dependence of $\sigma_{12}$ on $r_p$ may reflect the scaling of $\bar{v}_{12}$ more than it reflects the true behavior of $\sigma_{12}$.

## 5 RESULTS

### *5.1* $\xi(r_p, \pi)$ *for the Data*

Figures 9 and 10 show contours of $\xi(r_p, \pi)$ for CfA2N, CfA2S, SSRS2 and CfA2+SSRS2. In order to facilitate comparisons with other surveys and to depict more clearly the distortions in $\xi(r_p, \pi)$, we display $\xi(r_p, \pi)$ with linear $1h^{-1}$ Mpc binning. We emphasize that this binning is used for display purposes only; for model fitting, we use logarithmic binning exclusively.

Pairs are MV weighted in Figure 9 and uniformly weighted in Figure 10. The choice of weighting significantly affects $\xi(r_p, \pi)$. Although the sensitivity to weighting is small in CfA2+SSRS2, the remaining fluctuations emphasize that surveys on this scale do not probe a "fair" volume of the universe. The fluctuations in $\xi(r_p, \pi)$ with weighting are not surprising given the redshift distributions of these samples (Geller and Huchra 1989, Park *et al.* 1994, da Costa *et al.* 1994). In CfA2N, uniform weighting accentuates the generally low-density features nearby. Uniform weighting also increases the contribution of faint galaxies, which are less clustered than brighter galaxies (Giovanelli & Haynes 1989,Vogeley *et al.* 1994). On the other hand, MV weighting enhances the contribution of the Great Wall, which contains most of the groups and clusters in CfA2N. The clustering amplitude is consequently larger for MV weighting , and the orientation of this huge structure elongates $\xi(r_p, \pi)$ in the $r_p$ direction.

CfA2S reveals a similar trend. In this sample, the Perseus-Pisces wall at $\sim 5,000\,\mathrm{km\,s^{-1}}$ lies near the peak of the distribution $\phi(r)V(r)$ where $\phi$ is the selection function and $V$ is the volume of the shell at distance $r$. In this case, MV weighting enhances the low-density regions in the foreground and background, lowering the amplitude of $\xi(r_p, \pi)$. The shape of $\xi(r_p, \pi)$ on small scales is only mildly sensitive to the weighting in CfA2S. Like the Great Wall in CfA2N, the Perseus-Pisces wall runs perpendicular to the line of sight; as expected from the redshift distribution, uniform weighting enhances correlations in the $r_p$ direction.

The SSRS2 is fairly insensitive to the weighting scheme on scales smaller than $\sim 10h^{-1}$ Mpc. This insensitivity is reasonable, as the most obvious structure crosses the survey at an angle to both the $r_p$ and $\pi$ axes. On larger scales, MV weighting strongly enhances line-of-sight correlations, again reflecting the small number of independent structures sampled in this volume. Although uniform weighting provides insight into the fluctuations in $\xi(r_p, \pi)$ within the sample, it is a somewhat unnatural and inefficient weighting scheme; we therefore drop the uniformly weighted CF from further consideration. We calculate the spatial CF and the pairwise dispersion using MV weighting exclusively.

### *5.2 Spatial Correlation Function*

Figure 11 shows power-law fits to $w(r_p)$ for each sample. We summarize these fits in Table 3, which lists the parameters of the spatial CF obtained from the inversion in Equation 12. As discussed in §3, we transform the data vector $w(r_p)$ to diagonalize the covariance matrix and calculate $\tilde{\chi}^2$ from Equation 20. A single power law fits $w(r_p)$ for each sample at separations smaller than 10 $h^{-1}$ Mpc. The correlation lengths are roughly consistent for CfA2N and SSRS2: $r_0 = 5.83 \pm 0.18$ and $5.08 \pm 0.23$, respectively. The slopes are also consistent: for CfA2N, $\gamma = -1.80 \pm 0.03$; for the SSRS2, $\gamma = -1.85 \pm 0.06$. In CfA2S, the amplitude of the clustering is significantly weaker and the slope is steeper: $r_0 = 4.75 \pm 0.20 h^{-1}$ Mpc and $\gamma = -1.99 \pm 0.05$. Because $r_0$ and $\gamma$ are strongly correlated, these differences are not as large as they appear; if we constrain $\gamma$ to be -1.8, $r_0 = 5.4$ for CfA2S. For consistency, we use the fitted values for both the amplitude *and* the slope to calculate $\sigma_{12}$.

### *5.3 Pairwise Dispersion for CfA2 and SSRS2*

Figures 12-15 show cuts of $\xi(r_p, \pi)$ at each $r_p$ for CfA2N, CfA2S, SSRS2 and CfA2+SSRS2 respectively. We include fits of Equation 8 based on the two distributions $P(v_{12}^{\alpha})$ in Equations 15 and 18. Dashed lines represent the single-galaxy exponential (Equation 18); solid lines represent the pairwise exponential distribution. Again, we neglect the possible contribution of the mean pairwise infall $\bar{v}_{12}$.

Error bars in Figures 12-15 represent the raw bootstrap variances. Tables 4-7 list estimates of $\sigma_{12}$ at each value of $r_p$ for each sample. Columns 3-5 show fits to the single-galaxy exponential in Equation 18. Columns 6-8 give the fits for a pairwise exponential. Tables 4-7 also include fits using the infall model in Equation 19 with F=1.

In order to facilitate comparisons with models and with other surveys, we compute a best estimate for $\sigma_{12}$ as the mean of the first three bins in $r_p$. This mean, $\sigma_{12}(r_p \leq 0.8 h^{-1}\,\mathrm{Mpc})$, is listed for each sample in Table 8 along with its associated error. The variation in $\sigma_{12}(\leq 0.8)$ among samples is striking. For the exponential $P(v_{12}^{\alpha})$ with no infall, $\sigma_{12}(\leq 0.8) = 647 \pm 52\,\mathrm{km\,s^{-1}}$, $367 \pm 38\,\mathrm{km\,s^{-1}}$ and $272 \pm 42\,\mathrm{km\,s^{-1}}$ for CfA2N, CfA2S and SSRS2, respectively. In general, including infall increases the estimate of $\sigma_{12}(\leq 0.8)$, consistent with DP83.

The behavior of $\sigma_{12}$ as a function of $r_p$ varies with the infall model. Figure 16 shows $\sigma_{12}(r_p)$ for each sample. If we include infall, we find that $\sigma_{12}$ generally rises slowly or is effectively constant with $r_p$, consistent with the observations of DP83 and Mo *et al.* (1993). However, if we ignore infall, the dispersion drops with $r_p$ in each case. Because of the difficulties with the infall model at large $r_p$ highlighted in §4, we suggest that the dependence of $\sigma_{12}$ on $r_p$ measured with this technique should be interpreted with caution.

Using the individual samples with the standard binning, the errors preclude a strong rejection of either model for the velocity distribution function. In order to increase the signal-to-noise at small $r_p$, we recompute $\xi(r_p, \pi)$ for the combined CfA2+SSRS2 using linear, $1 h^{-1}$ Mpc bins (see figure 9). Figure 17 shows the cut in $\xi(r_p, \pi)$ at $r_p \leq 1 h^{-1}$ Mpc along with fits based on the velocity distributions in Equations 15 and 18. The single-galaxy exponential, represented by the dashed line, is clearly incompatible with the data ($\chi^2/\mathrm{d.o.f.} = 37.7/13, Q = 2 \times 10^{-4}$). The single-galaxy exponential is too flat at small velocities and drops too quickly at large velocities. On the other hand, the pairwise expo-nential reproduces the data well ($\chi^2$/d.o.f. $= 10.5/13, Q = 0.65$), and the corresponding dispersion 565±33 is consistent with our best estimate derived above, $\sigma_{12}(\leq 0.8) = 540\pm36$. Thus the velocity field in the nearest $\sim 120h^{-1}$ Mpc is remarkably consistent with an isotropic exponential distribution of *pairwise* velocities, consistent with previous surveys ( *e.g.*, DP83, Mo *et al.* 1993, Fisher *et al.* 1994b).

Table 8 summarizes the most important results of this analysis: the pairwise dispersions $\sigma_{12}(\leq 0.8)$ measured from the CfA2 and SSRS2 samples. For the combined sample, we estimate the error in $\sigma_{12}$ from the scatter in the individual sample estimates. This error (labeled "external") is simply the weighted sample standard deviation. Two points are clear: 1) $\sigma_{12}$ varies broadly among volumes $\leq 10^6\, h^{-3}\,\mathrm{Mpc}^3$, and 2) $\sigma_{12}$ for CfA2+SSRS2 significantly exceeds previous estimates of the pairwise velocity dispersion, which yield $\sigma_{12} \sim 300$. The agreement on $\sigma_{12}$ between Fisher *et al.* (1994b) and DP83 may be fortuitous; CfA1 fully samples only one rich cluster, and the IRAS survey is dominated by spiral galaxies, which avoid the cores of clusters. Indeed, Mo *et al.* (1993) and Zurek *et al.* (1994) find that the value of $\sigma_{12}$ derived from CfA1 varies significantly depending on how the velocity field around the Virgo cluster is treated. With no corrections to the velocity field, the estimate of $\sigma_{12}$ for CfA1 derived by Zurek *et al.* (1994) is consistent with $\sigma_{12}$ derived from CfA2+SSRS2. Furthermore, $\sigma_{12}$ for ellipticals and S0's is significantly larger than it is for spirals (Mo *et al.* 1993, Marzke *et al.* 1994b). Thus the variations in the observed $\sigma_{12}$ appear to stem from the contributions of a few rich clusters. We test this hypothesis in the following sections.

### *5.4 The Cluster Contribution*

To guage the contribution to $\sigma_{12}$ from rich clusters, we mask out galaxies within three degrees of Abell clusters with richness class $R \geq 1$. Figure 18 depicts $\xi(r_p, \pi)$ for these pruned samples. Not surprisingly, the removal of clusters suppresses the extension of $\xi(r_p, \pi)$ along the $\pi$ direction.

Table 3 includes fits to $w(r_p)$ with clusters removed. Except in CfA2S, the slope of the spatial CF does not change when we remove clusters, but the amplitude decreases significantly. In CfA2S, removing clusters increases the amplitude and flattens the slope. These changes in $\xi(r)$ have little effect on the calculation of $\sigma_{12}$ on small scales.

Tables 4-7 list $\sigma_{12}$ derived from the pruned samples. In most cases, $\sigma_{12}$ drops significantly when we remove clusters. The change is most dramatic in CfA2N, where $\sigma_{12}(\leq 0.8)$ drops from $647 \pm 52$ to $313 \pm 39$. Excluding all galaxies within $2500\,\mathrm{km\,s^{-1}}$ in CfA2N does not change this result significantly; in this case, $\sigma_{12} = 327 \pm 41$. Thus the infall pattern around the Virgo cluster does not affect our derivation of $\sigma_{12}$. The cores of rich clusters are the dominant source of variation in $\sigma_{12}$.

In CfA2S, we measure a similar decline in $\sigma_{12}$ from $367 \pm 38\,\mathrm{km\,s^{-1}}$ to $264 \pm 38\,\mathrm{km\,s^{-1}}$ when we remove clusters. On the other hand, the pruned SSRS2 sample is consistent with the SSRS2 as a whole. Because the SSRS2 contains no very rich clusters like Coma, this insensitivity is not surprising (Ramella *et al.* 1995). For the combined sample, we find $\sigma_{12}(\leq 0.8) = 295 \pm 31\,\mathrm{km\,s^{-1}}$ after removing clusters. Thus the difference between CfA2+SSRS2 and DP83 can be attributed entirely to the presence of rich clusters in CfA2.

## 6 DISCUSSION

### *6.1 Comparison with Cosmological Models*

The pairwise velocity dispersion measured by DP83 has stood for a decade as one of the strongest arguments against the standard CDM paradigm. Taken at face value, our results appear to alleviate the conflict. Figure 19 displays $\sigma_{12}$ measured from CfA2+SSRS2 along with a compendium of model predictions. We include only variants of CDM; we do not consider possible mixtures of CDM with other species, which provide an equally good match to the measured power spectrum (Summers *et al.* 1994, Klypin *et al.* 1993). Although unbiased flat CDM still fails to match the data, biased $\Omega = 1$ models are no longer ruled out by the pairwise velocity dispersion.

Our data also appear consistent with the open models discussed by Davis *et al.* (1985). Interestingly, open models seem to provide the best match to a wealth of observational constraints: the measured power spectrum, void probability function, and small-scale velocity field are all consistent with $\Omega h \approx 0.2$ CDM (Vogeley *et al.* 1992,1994,Park *et al.* 1994, Fisher *et al.* 1992). On the other hand, biased CDM, which satisfies our constraint from $\sigma_{12}$, does not reproduce the observed void probability function (Vogeley *et al.* 1995).

Unfortunately, the interpretation of $\sigma_{12}$ is complicated by the presence of rich clusters. Because we and others have shown that $\sigma_{12}$ is very sensitive to the exclusion of clusters, the pairwise dispersion we measure cannot be construed as the true "field" velocity dispersion. The open squares in Figure 18 show $\sigma_{12}$ from CfA2+SSRS2 after removing Abell clusters with $R \geq 1$. If the model predictions represent the *field*, then the constraints on all variants of CDM still stand.

Clearly, the definition of the "field" poses a significant problem for the interpretation of the small-scale velocities in both the models and in the real universe. For example, Couchman and Carlberg (1992) find that $\sigma_{12}$ varies between 100 $\mathrm{km\,s^{-1}}$ and 1000 $\mathrm{km\,s^{-1}}$ even in moderately dense regions. The value for $\sigma_{12}$ quoted in Couchman and Carlberg (1992) and reproduced in Figure 19 does not include the richest cluster in their simulation. With the current limitations on both observations and theory, the effort to constrain cosmological models demands careful, consistent filtering of both simulations and data to establish fair grounds for comparison.

In any clustering scheme, gravity builds up a continuum of bound systems. The removal of clusters is somewhat arbitrary, particularly when the surrounding large-scale structure is complex and the observed (or simulated) volumes are small. Ideally, one would like the simulations to reproduce $\sigma_{12}$ averaged over *all* galaxies. On the theoretical side, this demand calls for a detailed treatment of the interaction among galaxies in a large enough volume to include an ensemble of large, dense structures. Observationally, the requirement is a fair sample of the universe.

These are tall orders. In order to size up the task before us, we ask the following questions : 1) how does the distribution of cluster velocity dispersions affect $\sigma_{12}$ and 2) on what scale does $\sigma_{12}$ converge to its global mean?

### *6.2 Pairwise Dispersion from Virialized Systems*

We develop a simple model to describe the variation of $\sigma_{12}$ with survey volume. Redshift surveys indicate that a substantial fraction of galaxies aggregate into bound systems with velocity dispersions ranging from $\sim 100\,\mathrm{km\,s^{-1}}$ in loose groups to more than $1000\,\mathrm{km\,s^{-1}}$ in rich clusters (Geller and Huchra 1983, Nolthenius and White 1987, Ramella *et al.* 1989, Moore *et al.* 1993, Zabludoff *et al.* 1993). Because we know from §5.4 that dense systems contribute significantly to the field velocity dispersion $\sigma_{12}$, we can predict $\sigma_{12}$ and its variance in a given volume from the distribution of group and cluster velocity dispersions, $n(\sigma)$ (cf. Carlberg 1994). We can obtain $n(\sigma)$ either from the measured distributions (Zabludoff *et al.* 1993, Frenk *et al.* 1991) or directly from the power spectrum of fluctuations using the Press-Schechter formulation for the evolution of the mass spectrum (Press and Schechter 1974).

By definition, the statistic $\sigma_{12}$ is pair-weighted. Because the number of pairs in a cluster is proportional to the square of the number density of galaxies within the cluster, $\rho$, the mean pairwise dispersion is

$$\sigma_{12}^2 = 2\frac{\int_{\sigma_{min}}^{\sigma_{max}} \sigma^2 \rho^2(\sigma) n(\sigma)\, d\sigma}{\int_{\sigma_{min}}^{\sigma_{max}} \rho^2(\sigma) n(\sigma)\, d\sigma} \tag{21}$$

where $\sigma_{min}$ and $\sigma_{max}$ limit the range of group and cluster velocity dispersions under consideration. The largest measured dispersion is $\sigma = 1436\,\mathrm{km\,s^{-1}}$ (Oegerle 1994, private communication); extrapolation of $n(\sigma)$ to larger values is uncertain. We assume in Equation 21 that both members of a pair inhabit the same system, a reasonable assumption on scales $\leq 1h^{-1}\,\mathrm{Mpc}$. Dropping this assumption increases the expected $\sigma_{12}$. The relation between the number density of galaxies within a single cluster and the cluster velocity dispersion, $\rho(\sigma)$, requires an assumption about the distribution function. We parameterize this dependence as $\rho \propto \sigma^m$:

$$\sigma_{12}^2 = 2\frac{\int_{\sigma_{min}}^{\sigma_{max}} \sigma^{2+2m} n(\sigma)\, d\sigma}{\int_{\sigma_{min}}^{\sigma_{max}} \sigma^{2m} n(\sigma)\, d\sigma} \tag{22}$$

For an isothermal sphere, $m = 2$. Because observational constraints on $m$ are weak, we assume $m = 2$ throughout this analysis. If $m > 2$, both $\sigma_{12}$ and the predicted variance in $\sigma_{12}$ increase. We ignore the contribution of an unbound, true field population. Including a low-dispersion field would reduce the estimated $\sigma_{12}$ but would not contribute significantly to the variance in $\sigma_{12}$.

### 6.2.1 $\sigma_{12}$ *from the Observed Distribution of Velocity Dispersions*

Zabludoff *et al.* (1993) measure the distribution of velocity dispersions for dense systems. Their sample includes dense Abell clusters with richness class $R \geq 1$ as well as a sample of dense groups selected from the first four slices of CfA2(Geller & Huchra 1989,Huchra *et al.* 1994). The authors conclude that the number density of systems declines exponentially with velocity dispersion: $n(\sigma) \propto 10^{-0.0015\sigma}$. Using a similar sample of groups from CfA1 (Huchra *et al.* 1983), Moore *et al.* 1993 find a steeper slope for $n(\sigma)$, roughly proportional to $10^{-0.004\sigma}$. Their comparison with the distribution of cluster velocity dispersions culled from the literature (Frenk *et al.* 1990) suggests a discontinuity in $n(\sigma)$ at $\sigma \sim 700\text{kms}^{-1}$, where the distribution becomes flatter. The authors attribute this flattening to field contamination in the cluster redshift distributions, which artificially inflate the measured dispersions. The $n(\sigma)$ inferred from the distribution of cluster X-ray temperatures falls somewhere in between the results of Zabludoff *et al.* (1993) and Moore *et al.* (1993); The X-ray temperature distributions follow a power law corresponding to $n(\sigma) \propto \sigma^{-5}$ over the observed range (Edge *et al.* 1990, Henry & Arnaud 1991).

We use these measurements of $n(\sigma)$ to estimate the volume-averaged $\sigma_{12}$ according to Equation 22. We show the range of $n(\sigma)$ from the observations as dashed lines in Figure 20. Following convention, we plot the cumulative distribution $n(\geq \sigma)$. Each distribution is normalized to the mean density of Abell clusters with $R \geq 1$ $n(\sigma > 700\,\text{km}\,\text{s}^{-1}) = 6.6 \times 10^{-6}\text{h}^3\text{Mpc}^{-3}$ (Zabludoff *et al.* 1993).

Figure 21 shows the expected mean pairwise dispersion,$\sigma_{12}$, as a function of $\sigma_{max}$ from Equation 22 for the observed range of $n(\sigma)$. The convergence of $\sigma_{12}$ depends on the shape of $n(\sigma)$ and on the extrapolation of the observed distributions beyond the observed range of $\sigma$. For steeper $n(\sigma)$, $\sigma_{12}$ converges to lower values and at lower $\sigma_{max}$. If we simply extrapolate the distribution $n(\sigma)$ derived by Zabludoff *et al.* (1993), $\sigma_{12}$ does not converge until $\sigma_{max} \sim 3500\,\text{km}\,\text{s}^{-1}$, well beyond the observed range.

The velocity dispersion $\sigma_{conv}$ at which $\sigma_{12}$ converges determines a minimum volume required for the statistic to be representative. In order to predict the variance of $\sigma_{12}$ for a given survey volume $V$, we perform a simple Monte-Carlo simulation. The number of clusters $N$ in the hypothetical sample volume $V$ is $V[n(>\sigma_{min}) - n(>\sigma_{max})]$. As in Zabludoff *et al.* (1993), we assume $\sigma_{min} = 100\,\text{km}\,\text{s}^{-1}$. The cutoff $\sigma_{max}$ is a free parameter. At each volume, we construct several Monte-Carlo realizations of a sample of $N$ clusters drawn from the distribution $n(\sigma)$ . We then measure $\sigma_{12} = \sum_{i=1}^{N} \sigma_i^{2m+2} / \sum_{i=1}^{N} \sigma_i^{2m}$. Because of the large weight given to high-dispersion systems, our analysis of the bias and variance in $\sigma_{12}$ is much more sensitive to $\sigma_{max}$ than it is to $\sigma_{min}$.

Figure 22 shows the distribution of the fractional error in $\sigma_{12}$,$\delta\sigma_{12}/\sigma_{12}$, over the range $V = 10^5 - 10^8 h^{-3}\text{Mpc}^3$ for the range of $n(\sigma)$ derived from observations. Arrows indicate the volumes of CfA1, CfA2N and CfA2+SSRS2. The solid line represents the median $\sigma_{12}$; dotted lines enclose 68% of the results for each volume. Because $\sigma_{12}$ is weighted by the square of the density, the shot noise caused by the one or two richest clusters in each sample dominates the variance at small volumes. The distribution of $\sigma_{12}$ is noticeably skew with a tail toward large $\sigma_{12}$. Furthermore, $\sigma_{12}$ is biased toward low values in samples with small volumes. As the volume increases, samples include significant numbers of clusters with $\sigma \sim \sigma_{conv}$, and the variance drops accordingly. On the scales of the surveys, the variance and bias are larger than the random errors in $\sigma_{12}$, especially for the shallow distribution in Zabludoff *et al.* (1993).

Figure 22 also shows the sensitivity to $\sigma_{max}$. Increasing $\sigma_{max}$ from 2000 $\mathrm{km\,s^{-1}}$ to 3000 $\mathrm{km\,s^{-1}}$ affects the convergence of $\sigma_{12}$ much less than changing the slope of $n(\sigma)$ over the observed range. As expected, extrapolating to larger $\sigma$ slows the convergence of $\sigma_{12}$ with volume. There is no reason to expect that the observed distributions $n(\sigma)$ extend beyond the observed range; in the next section, we investigate a more physical model for the extremes of $n(\sigma)$.

#### *6.2.2 $\sigma_{12}$ from the Press-Schechter $n(\sigma)$*

The mass distribution of virialized systems at any epoch is largely determined by the low-order moments of the initial density perturbations. Assuming spherical collapse of Gaussian perturbations, Press and Schechter (1974) describe the evolution of the distribution of mass clumps for various assumptions about the power spectrum of initial density perturbations. Although the ingredients of the model are approximate, N-body simulations of the non-linear (and non-spherical) collapse of structure show a remarkable agreement between the Press-Schechter prediction and the measured $n(M, z)$ for a wide range of initial conditions (Efstathiou *et al.* 1988, Carlberg and Couchman 1989). Recent attempts to improve the analytic model suggest that the low and high-mass ends of $n(M, z)$ may not be accurate (e.g. Peacock and Heavens 1990, Monaco 1994), but we are primarily interested in the regions in between.

To estimate the distribution of cluster velocity dispersions, we calculate the Press-Schechter distribution in the form presented by Carlberg *et al.* (1994):

$$n(\sigma)\,d\sigma = \frac{9c_v^3 H_0^3}{4\pi\sqrt{2\pi}} \frac{(1+z)^{3/2}}{\sigma^4} \frac{d\ln\Delta(M(r))}{d\ln\sigma} \nu e^{-\nu^2/2}\,d\sigma \tag{23}$$

where $\nu = 1.68(1+z)/(\sigma_8\Delta(M))$, $\sigma_8$ is the normalization of the mass fluctuation spectrum, and $\Delta(M(r))$ is the rms mass fluctuation on the scale $r$. We use the analytic fit to $\Delta(r)$ for flat CDM given by Narayan and White (1988). The constant $c_v$ defines the relation between $\sigma$ and mass M: $\sigma = c_v H_0 R(1+z)^{1/2}$, where $M = 4\pi\rho_0 R^3/3$. White and Frenk (1991) measure $c_v = 1.18$ for the spherical collapse of a tophat sphere. For standard CDM, the only free parameter is the normalization, $\sigma_8$.

Solid lines in Figure 20 show the cumulative distribution of velocity dispersions given by the Press-Schechter prescription for two normalizations of the power spectrum. The normalization $\sigma_8 = 1.0$ corresponds approximately to the measured quadrupole anisotropy of the microwave background (Wright *et al.* 1992). The second normalization, $\sigma_8 = 0.5$, indicates the low end of the range currently favored by models of biased galaxy formation (Couchman and Carlberg 1992, Cen and Ostriker 1993). The distribution of Abell cluster velocity dispersions (Zabludoff *et al.* 1993) is significantly shallower than the CDM predic-tions for any acceptable value of $\sigma_8$ (this discrepancy is unlikely to be caused entirely by projection effects; c.f. Zabludoff et al 1990 and Frenk *et al.* 1990).

Figure 21 shows the convergence of $\sigma_{12}$ computed from the models. Even for $\sigma_8 = 1.0$, $\sigma_{12}$ converges within the observed range of cluster velocity dispersions. $\sigma_{12}$ converges to 1239 $\mathrm{km\,s^{-1}}$ for COBE-normalized CDM, remarkably consistent with the dark-matter velocity dispersion derived by Cen and Ostriker (1993) for $\sim 1h^{-1}$ Mpc pair separations.

Figure 23 shows $\delta\sigma_{12}/\sigma_{12}$ given by the Monte-Carlo procedure described above but using $n(\sigma)$ from the models. As expected, the convergence of $\sigma_{12}$ with volume depends strongly on $\sigma_8$. For $\sigma_8 = 0.5$, a 10% measurement of $\sigma_{12}$ is obtained in a survey volume $\sim 10^5\mathrm{Mpc}^3$. For unbiased CDM, the required survey volume is much larger, approximately $5 \times 10^6\mathrm{Mpc}^3$, which is larger than any existing redshift survey but comparable with the planned Sloan Digital Sky Survey (Gunn and Weinberg 1995).

We conclude that the variation in $\sigma_{12}$ among subsamples of CfA2+SSRS2 is consistent with expectations based on the observed distribution of group and cluster velocity dispersions. The volume required for a 10% measurement of $\sigma_{12}$ is larger than the volume surveyed by CfA2+SSRS2 in an unbiased, flat CDM universe. Furthermore, we show that $\sigma_{12}$ is biased low by approximately 20% on the scale of CfA1 (used by DP83) for the same model. The bias and variance are worse if $n(\sigma)$ observed by Zabludoff *et al.* (1993) extends to larger $\sigma$. Convergence occurs on a smaller scale for high-bias CDM models. Our model is admittedly approximate; cluster-cluster correlations increase the expected variance, while velocity bias tends to decrease it. It is clear, however, that large redshift surveys as well as large, high-resolution simulations are required for a robust comparison between observed and predicted pairwise velocity dispersions.

## 7 CONCLUSIONS

We combine the CfA Redshift Survey (CfA2) and the Southern Sky Redshift Survey (SSRS2) to estimate the pairwise velocity dispersion of galaxies on a scale of $\sim 1h^{-1}$ Mpc. Both surveys are complete to an apparent magnitude limit $B(0) = 15.5$. Our sample includes 12,812 galaxies distributed in a volume $1.8 \times 10^6\, h^{-3}\,\mathrm{Mpc}^3$. We conclude:

1) The pairwise velocity dispersion of galaxies in the combined CfA2+SSRS2 redshift survey is $\sigma_{12} = 540\,\mathrm{km\,s^{-1}} \pm 180\,\mathrm{km\,s^{-1}}$. Both the estimate and the error in $\sigma_{12}$ exceed the canonical values $\sigma_{12} = 340 \pm 40$ measured by Davis and Peebles (1983) using CfA1.

2) We derive the uncertainty in $\sigma_{12}$ from the variation among subsamples with volumes on the order of $7 \times 10^5\ h^{-3}$ Mpc$^3$. This variation is nearly an order of magnitude larger than the formal error 36 $\mathrm{km\,s^{-1}}$ derived from least-squares fits to the CfA2+SSRS2 correlation function. This variation among samples is consistent with the conclusions of Mo *et al.* (1993) for a number of smaller surveys and with the analysis of CfA1 by Zurek *et al.* (1994).

3) When we remove Abell clusters with $R \geq 1$ from our sample, the pairwise velocity dispersion of the remaining galaxies drops to $295\,\mathrm{km\,s^{-1}} \pm 99\,\mathrm{km\,s^{-1}}$. This dispersion agrees with the results for CfA1 (DP83) and with the results for the IRAS 1.2 Jy survey (Fisher *et al.* 1994b), both of which are biased against rich clusters.

4) The dominant source of variance in $\sigma_{12}$ is the shot noise contributed by dense virialized systems. Because $\sigma_{12}$ is pair-weighted, the statistic is sensitive to the few richest systems in the volume. This sensitivity has two consequences. First, $\sigma_{12}$ is biased low in small volumes, where the number of clusters is small. Second, we can estimate the variance in $\sigma_{12}$ as a function of survey volume from the distribution of cluster and group velocity dispersions $n(\sigma)$. For either a COBE-normalized CDM universe or for the observed distribution of Abell cluster velocity dispersions, the volume required for $\sigma_{12}$ to converge ($\delta\sigma_{12}/\sigma_{12} < 0.1$) is $\sim 5 \times 10^6 h^{-3}$ Mpc$^3$, larger than the volume of CfA2+SSRS2.

5) The distribution of *pairwise* velocities is consistent with an isotropic exponential with velocity dispersion independent of scale. The inferred *single*-galaxy velocity distribution function is incompatible with an isotropic exponential. Thus the observed kinematics of galaxies differ from those measured in the hydrodynamic simulations of Cen and Ostriker (1993). On the other hand, our observations appear to be consistent with the distribution function measured by Zurek *et al.* (1994) using collisionless simulations with much higher resolution than Cen & Ostriker (1993). The large dynamic range in the simulations by Zurek *et al.* (1994) affords an accurate treatment of galaxy interactions, which dramatically alter the dynamical evolution the galaxy distribution (Couchman and Carlberg 1992,Zurek *et al.* 1994). The agreement between the observed velocity distribution and the one predicted by these high-resolution simulations may be another clue that mergers play an important role in the evolution of galaxies and large-scale structure.

We thank Mary Crone and Gus Evrard for generously providing one of their N-body simulations for the tests in §4. We also thank the referee, Jim Peebles, for many useful suggestions which improved the paper substantially. We thank Mike Vogeley, Ann

Zabludoff, Dan Lebach, Eric Blackman and Karl Fisher for discussions which contributed significantly to this paper. This work is supported in part by NASA grant NAGW-201 and by the Smithsonian Institution. R.M. acknowledges NASA grant NGT-50819.

## TABLE 1

PREVIOUS MEASUREMENTS OF $\sigma_{12}$

| Authors | Sample | $N_{gal}$ | $\sigma_{12}(\mathrm{km\,s^{-1}})$ |
|---|---|---|---|
| Geller & Peebles 1973 | RC1 | 226 | 206 |
| Peebles 1976 | RC1 | 422 | 200[a] |
| Peebles 1979 | KOS[b] | 166 | 500 |
| Davis & Peebles 1983 | CfA1[c] | 2397 | 340±40 |
| Bean *et al.* 1983 | AARS[d] | 320 | 250± 50 |
| Hale-Sutton *et al.* 1989 | SAAO[e] | 264 | 600 ± 140 |
| Mo *et al.* 1993 | Several | $\sim$ 700 –2400 | $\sim$ 200 –1000 |
| Fisher *et al.* 1994 | IRAS 1.2 Jy | 5313 | 317 ± 45 |
| Zurek *et al.* 1994 | CfA1 | 2397 | $\sim$ 300–600 |

[(a)]Excludes the Virgo cluster. Including Virgo, $\sigma_{12} \approx 600\,\mathrm{km\,s^{-1}}$.

[(b)]Kirshner, Oemler & Schechter 1978

[(c)]Huchra *et al.* 1983

[(d)]Peterson *et al.* 1983

[(e)]Metcalfe *et al.* 1989

**TABLE 2**

SURVEY SAMPLES

| Sample | $N_{gal}$ | $\Omega$ (ster) | Volume ($10^5\ h^{-3}\,\mathrm{Mpc}^3$) |
|---|---|---|---|
| CfA2 North | 6480 | 1.23 | 7.1 |
| CfA2 South | 2741 | 0.77 | 4.4 |
| SSRS2 | 3591 | 1.12 | 6.5 |
| CfA2+SSRS2 | 12812 | 3.12 | 18.0 |

## TABLE 3

CORRELATION FUNCTION $\xi(r)$

| Sample | $r_0\,(h^{-1}\,\mathrm{Mpc})$ | $\gamma$ | $\chi^2$/d.o.f. |
|---|---|---|---|
| CfA2 North | 5.83±0.18 | -1.80±0.03 | 11.5/18 |
| No Clusters | 4.77±0.17 | -1.80±0.04 | 16.8/18 |
| CfA2 South ($b \geq 10°$) | 4.75±0.20 | -1.99±0.05 | 18.9/18 |
| No Clusters | 3.74±0.18 | -2.11±0.08 | 12.0/18 |
| SSRS2 | 5.08±0.23 | -1.85±0.06 | 15.1/18 |
| No Clusters | 4.77±0.21 | -1.80±0.05 | 12.2/18 |
| CfA2+SSRS2 | 5.97±0.15 | -1.81±0.02 | 19.8/18 |
| No Clusters | 4.95±0.13 | -1.73±0.03 | 26.4/18 |

**TABLE 4**

PAIRWISE DISPERSION FOR CfA2 NORTH

| Comments | $r_p$ ($h^{-1}$ Mpc) | $\sigma_1^a$ (km s$^{-1}$) | $\chi^2$/dof | Q | $\sigma_{12}^b$ (km s$^{-1}$) | $\chi^2$/dof | Q |
|---|---|---|---|---|---|---|---|
| | 0.15 | 190±27 | 9.7/13 | 0.72 | 578±77 | 7.8/13 | 0.86 |
| | 0.3 | 259±27 | 5.2/13 | 0.97 | 592±88 | 4.6/13 | 0.98 |
| No Infall | 0.6 | 305±37 | 8.5/13 | 0.81 | 770±116 | 4.9/13 | 0.98 |
| | 1.2 | 274±26 | 10.1/13 | 0.69 | 697±92 | 5.0/13 | 0.98 |
| | 2.4 | 226±29 | 7.3/13 | 0.88 | 510±85 | 6.6/13 | 0.92 |
| | 4.8 | 79±37 | 7.4/13 | 0.88 | 149±89 | 7.5/13 | 0.88 |
| | 0.15 | 199±27 | 10.2/13 | 0.68 | 509±75 | 7.8/13 | 0.86 |
| | 0.3 | 270±26 | 5.3/13 | 0.97 | 640±80 | 4.6/13 | 0.98 |
| Infall | 0.6 | 329±35 | 8.8/13 | 0.79 | 825±124 | 4.8/13 | 0.98 |
| | 1.2 | 317±26 | 11.2/13 | 0.60 | 840±103 | 4.6/13 | 0.98 |
| | 2.4 | 307±26 | 8.4/13 | 0.82 | 751±93 | 6.9/13 | 0.91 |
| | 4.8 | 238±19 | 5.5/13 | 0.96 | 588±61 | 5.2/13 | 0.97 |
| | 0.15 | 182±27 | 2.2/13 | 1.00 | 384±82 | 2.4/13 | 1.00 |
| | 0.3 | 128±22 | 20.0/13 | 0.09 | 222±55 | 22.3/13 | 0.05 |
| No Infall | 0.6 | 157±26 | 4.6/13 | 0.98 | 333±74 | 6.1/13 | 0.94 |
| No Clusters | 1.2 | 128±21 | 6.6/13 | 0.92 | 276±57 | 7.3/13 | 0.89 |
| | 2.4 | 109±27 | 4.4/13 | 0.99 | 219±68 | 4.7/13 | 0.98 |
| | 4.8 | 7±156 | 4.4/13 | 0.99 | 65±221 | 19.2/13 | 0.12 |
| | 0.15 | 189±27 | 2.3/13 | 1.00 | 407±104 | 2.4/13 | 1.00 |
| | 0.3 | 143±20 | 16.3/13 | 0.23 | 258±48 | 18.8/13 | 0.13 |
| Infall | 0.6 | 180±24 | 4.1/13 | 0.99 | 392±70 | 5.6/13 | 0.96 |
| No Clusters | 1.2 | 170±18 | 6.6/13 | 0.92 | 388±56 | 7.3/13 | 0.89 |
| | 2.4 | 191±19 | 3.0/13 | 1.00 | 435±61 | 3.5/13 | 1.00 |
| | 4.8 | 197±20 | 8.9/13 | 0.78 | 452±61 | 12.6/13 | 0.48 |

[a] $P(v_{12}^{\alpha})$ from Equation 18, assuming isotropic exponential $P(v_1)$.

[b] Isotropic exponential $P(v_{12}^{\alpha})$ (Equation 15).

**TABLE 5**

PAIRWISE DISPERSION FOR CfA2 SOUTH

| Comments | $r_p$ ($h^{-1}$ Mpc) | $\sigma_1^a$ (km s$^{-1}$) | $\chi^2$/dof | Q | $\sigma_{12}^b$ (km s$^{-1}$) | $\chi^2$/dof | Q |
|---|---|---|---|---|---|---|---|
| | 0.15 | 166±26 | 6.0/13 | 0.95 | 400±78 | 4.9/13 | 0.98 |
| No Infall | 0.3 | 198±35 | 5.0/13 | 0.98 | 457±119 | 3.7/13 | 0.99 |
| | 0.6 | 109±15 | 16.3/13 | 0.23 | 244±45 | 15.0/13 | 0.31 |
| | 1.2 | 147±23 | 8.9/13 | 0.78 | 348±69 | 5.0/13 | 0.97 |
| | 2.4 | 109±29 | 5.6/13 | 0.96 | 243±76 | 4.3/13 | 0.99 |
| | 4.8 | 63±85 | 5.6/13 | 0.96 | 115±197 | 5.7/13 | 0.96 |
| | 0.15 | 179±28 | 6.3/13 | 0.93 | 366±80 | 4.7/13 | 0.98 |
| | 0.3 | 207±34 | 5.0/13 | 0.97 | 461±104 | 3.6/13 | 1.00 |
| Infall | 0.6 | 131±14 | 16.9/13 | 0.21 | 295±45 | 14.4/13 | 0.34 |
| | 1.2 | 197±23 | 10.9/13 | 0.62 | 463±72 | 4.6/13 | 0.98 |
| | 2.4 | 199±24 | 8.9/13 | 0.78 | 483±80 | 3.7/13 | 0.99 |
| | 4.8 | 198±35 | 3.1/13 | 1.00 | 475±111 | 4.8/13 | 0.98 |
| | 0.15 | 130±22 | 0.4/8 | 1.00 | 254±65 | 1.0/8 | 1.00 |
| | 0.3 | 164±39 | 4.1/8 | 0.85 | 353±111 | 3.4/8 | 0.91 |
| No Infall | 0.6 | 92±19 | 1.9/8 | 0.98 | 185±53 | 2.1/8 | 0.98 |
| No Clusters | 1.2 | 78±24 | 0.4/8 | 1.00 | 162±61 | 0.4/8 | 1.00 |
| | 2.4 | 15±210 | 2.6/8 | 0.96 | 27±472 | 2.7/8 | 0.95 |
| | 4.8 | –±— | —/- | —- | –±— | —-/- | —- |
| | 0.15 | 135±21 | 1.1/8 | 0.99 | 264±52 | 1.0/8 | 1.00 |
| | 0.3 | 170±40 | 4.1/8 | 0.85 | 160±118 | 11.6/8 | 0.17 |
| Infall | 0.6 | 110±17 | 2.0/8 | 0.98 | 225±48 | 2.0/8 | 0.98 |
| No Clusters | 1.2 | 118±18 | 1.0/8 | 1.00 | 253±54 | 0.4/8 | 1.00 |
| | 2.4 | 108±22 | 4.7/8 | 0.78 | 256±70 | 3.6/8 | 0.89 |
| | 4.8 | 150±41 | 1.8/8 | 0.99 | 330±113 | 2.1/8 | 0.98 |

$^{(a)}$ $P(v_{12}^{\alpha})$ from Equation 18, assuming isotropic exponential $P(v_1)$.

$^{(b)}$ Isotropic exponential $P(v_{12}^{\alpha})$ (Equation 15).

## TABLE 6

PAIRWISE DISPERSION FOR SSRS2

| Comments | $r_p$ ($h^{-1}$ Mpc) | $\sigma_1^a$ (km s$^{-1}$) | $\chi^2$/dof | Q | $\sigma_{12}^b$ (km s$^{-1}$) | $\chi^2$/dof | Q |
|---|---|---|---|---|---|---|---|
| | 0.15 | 150±32 | 3.4/13 | 1.00 | 329±70 | 3.3/13 | 1.00 |
| No Infall | 0.3 | 105±30 | 10.2/13 | 0.68 | 199±86 | 11.3/13 | 0.59 |
| | 0.6 | 131±23 | 10.2/13 | 0.68 | 287±66 | 5.1/13 | 0.97 |
| | 1.2 | 109±28 | 4.7/13 | 0.98 | 210±73 | 5.8/13 | 0.95 |
| | 2.4 | 48±45 | 3.8/13 | 0.99 | 95±98 | 3.8/13 | 0.99 |
| | 4.8 | 11±151 | 13.0/13 | 0.45 | 9±484 | 20.9/13 | 0.08 |
| | 0.15 | 160±29 | 3.6/13 | 0.99 | 334±95 | 3.3/13 | 1.00 |
| | 0.3 | 115±26 | 8.2/13 | 0.83 | 210±69 | 9.2/13 | 0.76 |
| Infall | 0.6 | 155±21 | 5.4/13 | 0.96 | 345±65 | 5.2/13 | 0.97 |
| | 1.2 | 152±22 | 3.8/13 | 0.99 | 324±65 | 5.5/13 | 0.96 |
| | 2.4 | 156±21 | 7.8/13 | 0.86 | 367±62 | 5.2/13 | 0.97 |
| | 4.8 | 189±28 | 9.0/13 | 0.78 | 453±80 | 7.1/13 | 0.89 |
| | 0.15 | 120±32 | 3.1/8 | 0.93 | 227±72 | 3.7/8 | 0.88 |
| | 0.3 | 95±25 | 3.9/8 | 0.86 | 179±67 | 4.6/8 | 0.80 |
| No Infall | 0.6 | 109±23 | 3.8/8 | 0.87 | 226±65 | 4.5/8 | 0.81 |
| No Clusters | 1.2 | 82±33 | 2.2/8 | 0.97 | 161±65 | 2.3/8 | 0.97 |
| | 2.4 | 43±213 | 3.4/8 | 0.91 | 85±59 | 2.3/8 | 0.97 |
| | 4.8 | 4±158 | 3.8/8 | 0.87 | 9±64 | 4.6/8 | 0.80 |
| | 0.15 | 131±27 | 2.8/8 | 0.94 | 241±64 | 3.4/8 | 0.91 |
| | 0.3 | 109±22 | 3.1/8 | 0.93 | 213±61 | 3.7/8 | 0.88 |
| Infall | 0.6 | 131±20 | 3.9/8 | 0.87 | 283±63 | 4.5/8 | 0.81 |
| No Clusters | 1.2 | 133±23 | 1.8/8 | 0.99 | 275±64 | 1.9/8 | 0.98 |
| | 2.4 | 142±20 | 5.9/8 | 0.66 | 317±59 | 4.4/8 | 0.82 |
| | 4.8 | 126±23 | 5.0/8 | 0.75 | 277±64 | 4.9/8 | 0.77 |

[(a)] $P(v_{12}^\alpha)$ from Equation 18, assuming isotropic exponential $P(v_1)$.

[(b)] Isotropic exponential $P(v_{12}^\alpha)$ (Equation 15).

**TABLE 7**

PAIRWISE DISPERSION FOR CfA2+SSRS2

| Comments | $r_p$ ($h^{-1}$ Mpc) | $\sigma_1^a$ (km s$^{-1}$) | $\chi^2$/dof | Q | $\sigma_{12}^b$ (km s$^{-1}$) | $\chi^2$/dof | Q |
|---|---|---|---|---|---|---|---|
| No Infall | 0.15 | 233±22 | 4.9/13 | 0.98 | 545±65 | 1.4/13 | 1.00 |
| | 0.3 | 224±21 | 9.2/13 | 0.76 | 540±57 | 5.6/13 | 0.96 |
| | 0.6 | 174±16 | 28.5/13 | 0.01 | 534±68 | 16.1/13 | 0.24 |
| | 1.2 | 209±18 | 13.8/13 | 0.39 | 515±55 | 5.7/13 | 0.96 |
| | 2.4 | 172±20 | 14.0/13 | 0.38 | 398±54 | 8.5/13 | 0.81 |
| | 4.8 | 9±231 | 14.0/13 | 0.38 | 111±93 | 4.7/13 | 0.98 |
| Infall | 0.15 | 239±19 | 5.1/13 | 0.97 | 564±65 | 1.5/13 | 1.00 |
| | 0.3 | 238±21 | 9.8/13 | 0.71 | 579±71 | 5.6/13 | 0.96 |
| | 0.6 | 200±16 | 32.7/13 | 0.01 | 599±64 | 15.6/13 | 0.27 |
| | 1.2 | 260±17 | 17.0/13 | 0.20 | 647±57 | 5.0/13 | 0.97 |
| | 2.4 | 274±17 | 20.2/13 | 0.09 | 654±56 | 6.4/13 | 0.93 |
| | 4.8 | 248±16 | 7.8/13 | 0.86 | 601±51 | 4.6/13 | 0.98 |
| No Infall No Clusters | 0.15 | 168±26 | 3.0/13 | 1.00 | 345±77 | 3.3/13 | 1.00 |
| | 0.3 | 139±19 | 16.4/13 | 0.23 | 269±48 | 19.0/13 | 0.12 |
| | 0.6 | 128±17 | 8.2/13 | 0.83 | 270±48 | 9.0/13 | 0.78 |
| | 1.2 | 109±14 | 9.1/13 | 0.77 | 230±39 | 9.7/13 | 0.72 |
| | 2.4 | 75±21 | 5.0/13 | 0.98 | 156±50 | 4.9/13 | 0.98 |
| | 4.8 | 9±401 | 5.0/13 | 0.98 | 13±528 | 62.2/13 | 0.00 |
| Infall No Clusters | 0.15 | 176±23 | 3.2/13 | 0.99 | 380±62 | 0.1/13 | 1.00 |
| | 0.3 | 158±17 | 13.2/13 | 0.43 | 304±43 | 16.3/13 | 0.23 |
| | 0.6 | 156±15 | 8.1/13 | 0.83 | 337±45 | 8.2/13 | 0.83 |
| | 1.2 | 153±12 | 8.5/13 | 0.81 | 351±38 | 8.1/13 | 0.83 |
| | 2.4 | 181±13 | 7.8/13 | 0.86 | 418±39 | 3.4/13 | 1.00 |
| | 4.8 | 190±19 | 4.4/13 | 0.99 | 444±56 | 6.5/13 | 0.93 |

[a] $P(v_{12}^{\alpha})$ from Equation 18, assuming isotropic exponential $P(v_1)$.

[b] Isotropic exponential $P(v_{12}^{\alpha})$ (Equation 15).

## TABLE 8

BEST ESTIMATES OF $\sigma_{12}$

| Sample | $\sigma_{12}(\leq 0.8)$ [$\mathrm{km\,s^{-1}}$] | Formal Error [$\mathrm{km\,s^{-1}}$] | External Error [$\mathrm{km\,s^{-1}}$] |
|---|---|---|---|
| CfA2 North | 647 | 52 | |
| CfA2 South | 367 | 38 | |
| SSRS2 | 272 | 42 | |
| CfA2+SSRS2 | 540 | 36 | 180 |

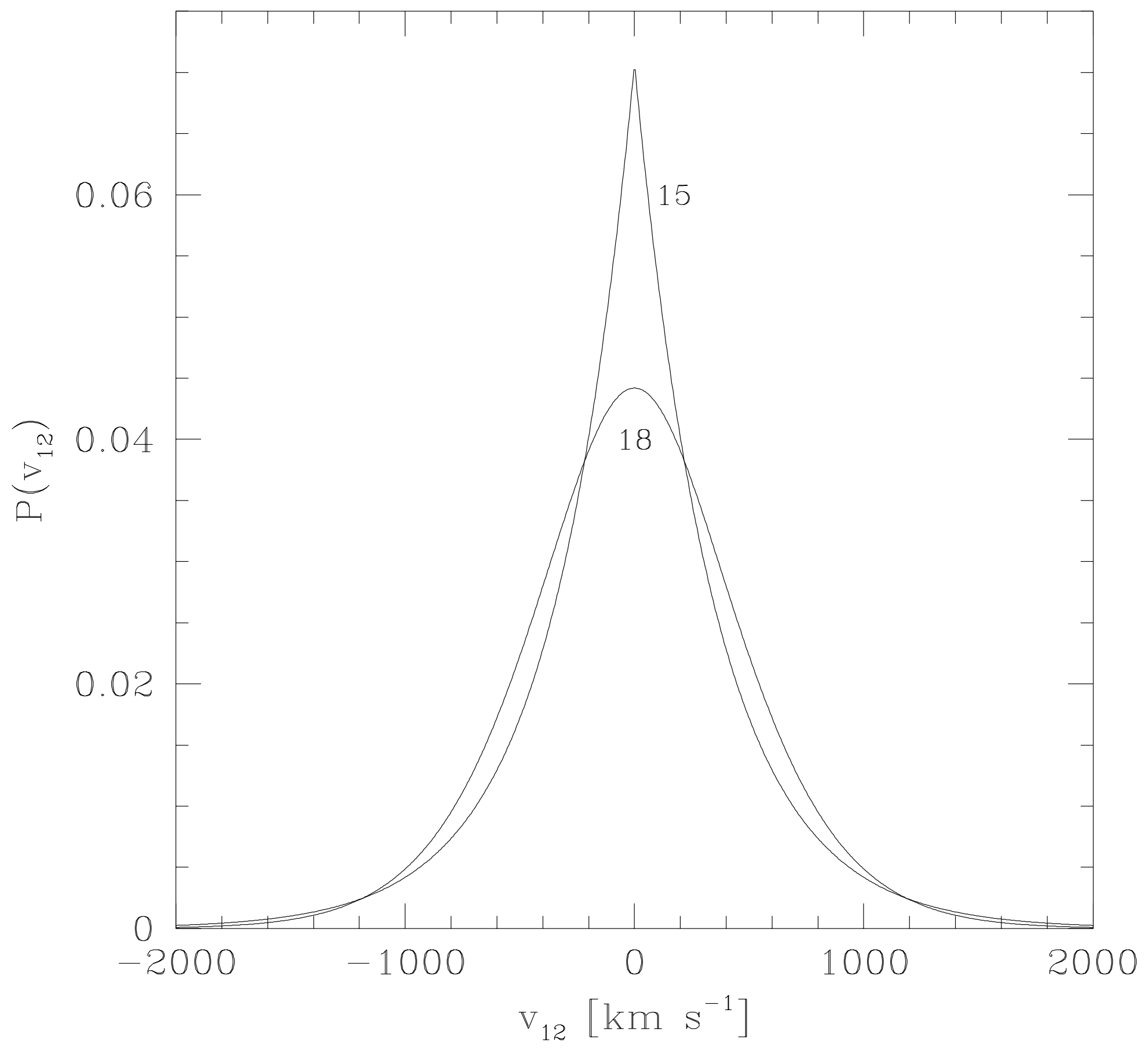

**Figure 1:** Pairwise velocity distribution functions from Equations 15 and 18. Curves are labeled by the equation number. Equation 15 describes the exponential pairwise distribution. Equation 18 describes the pairwise velocity distribution generated by an isotropic exponential distribution of single-galaxy velocities.

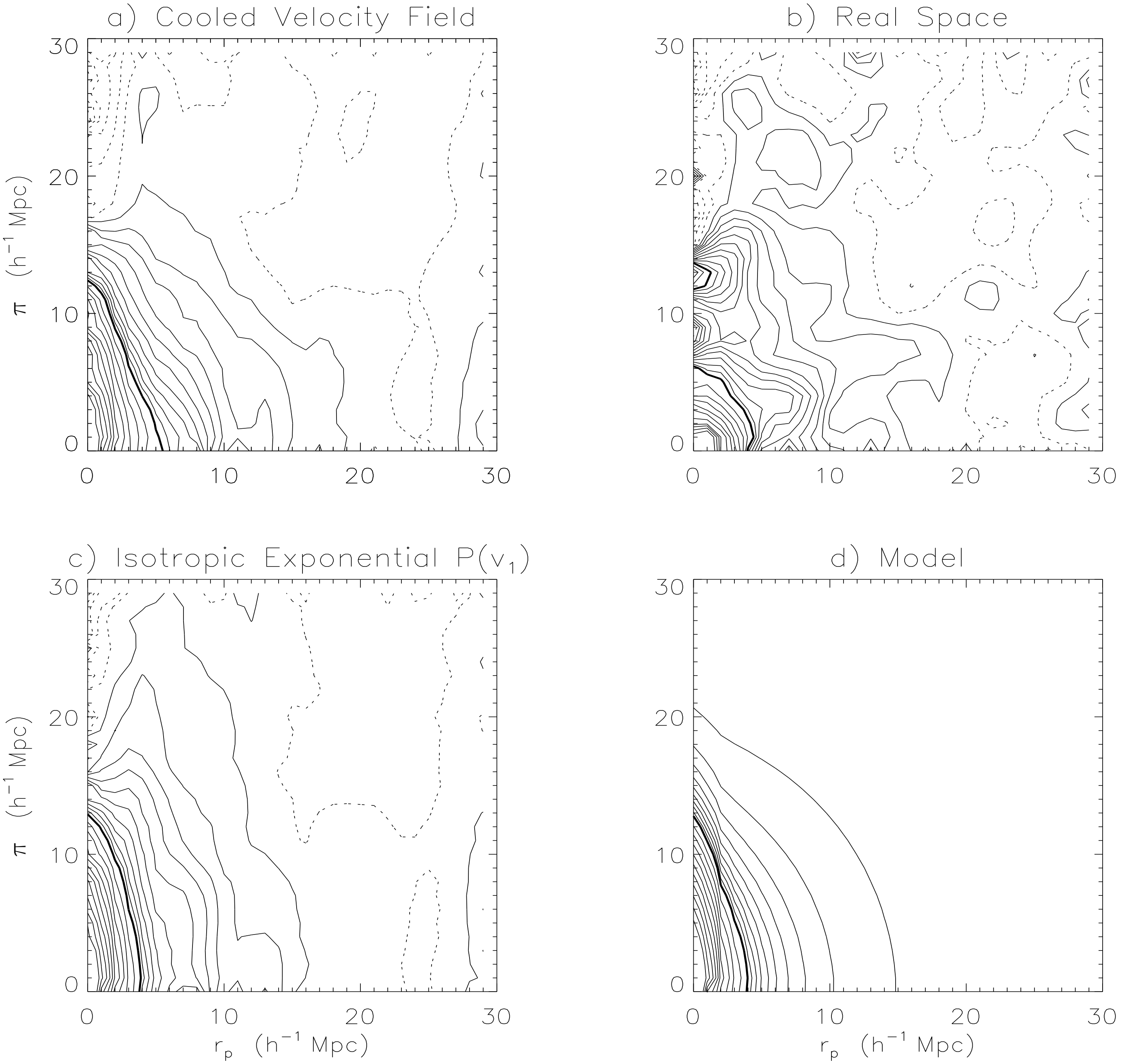

**Figure 2:** $\xi(r_p, \pi)$ for the N-body simulation. Solid contour indicates $\xi = 1$. Countour intervals are $\Delta\xi = 0.1$ for $\xi < 1$ and $\Delta \log \xi = 0.1$ for $\xi > 1$. Dotted contours indicate $\xi < 0$. (a) particle positions and velocities taken directly from the cooled simulation (see text), (b) pure Hubble flow - peculiar velocities set to zero, (c) particle velocities replaced with random deviates drawn from the isotropic exponential distribution in Equation 16, (d) best-fitting model from Equation 8.

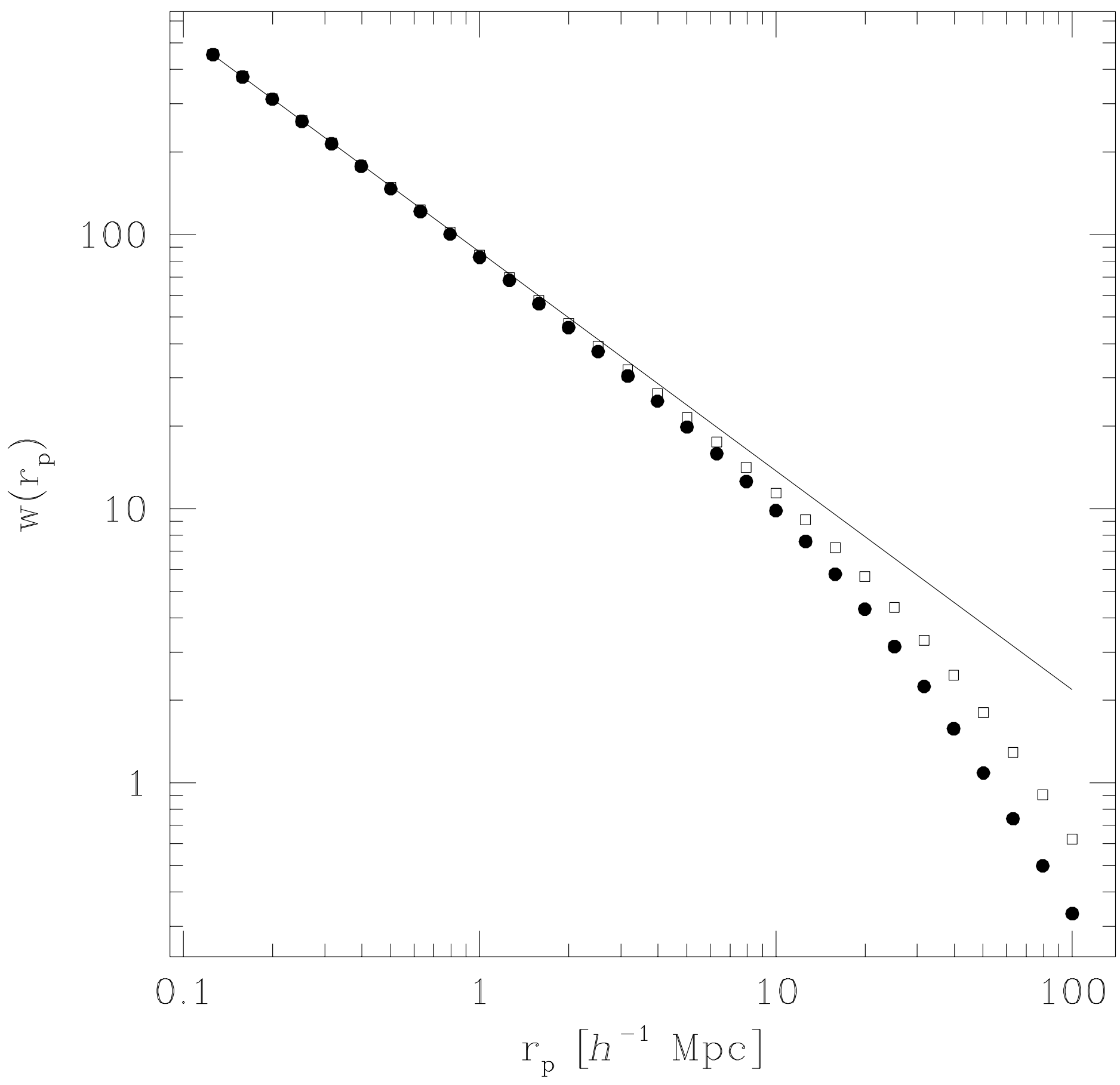

**Figure 3:** The projected CF, $w(r_p)$, determined from numerical integration of $\xi(r_p, \pi)$ out to $\pi_{max}$. The solid line is the input model: $r_0 = 5.8\,h^{-1}$ Mpc and $\gamma = -1.8$. $\xi(r_p, \pi)$ comes from Equation 8. Squares are for $\pi_{max} = 60\,h^{-1}$ Mpc, filled circles are for $\pi_{max} = 30\,h^{-1}$ Mpc. For the data, we integrate to $\pi_{max} = 30\,h^{-1}$ Mpc.

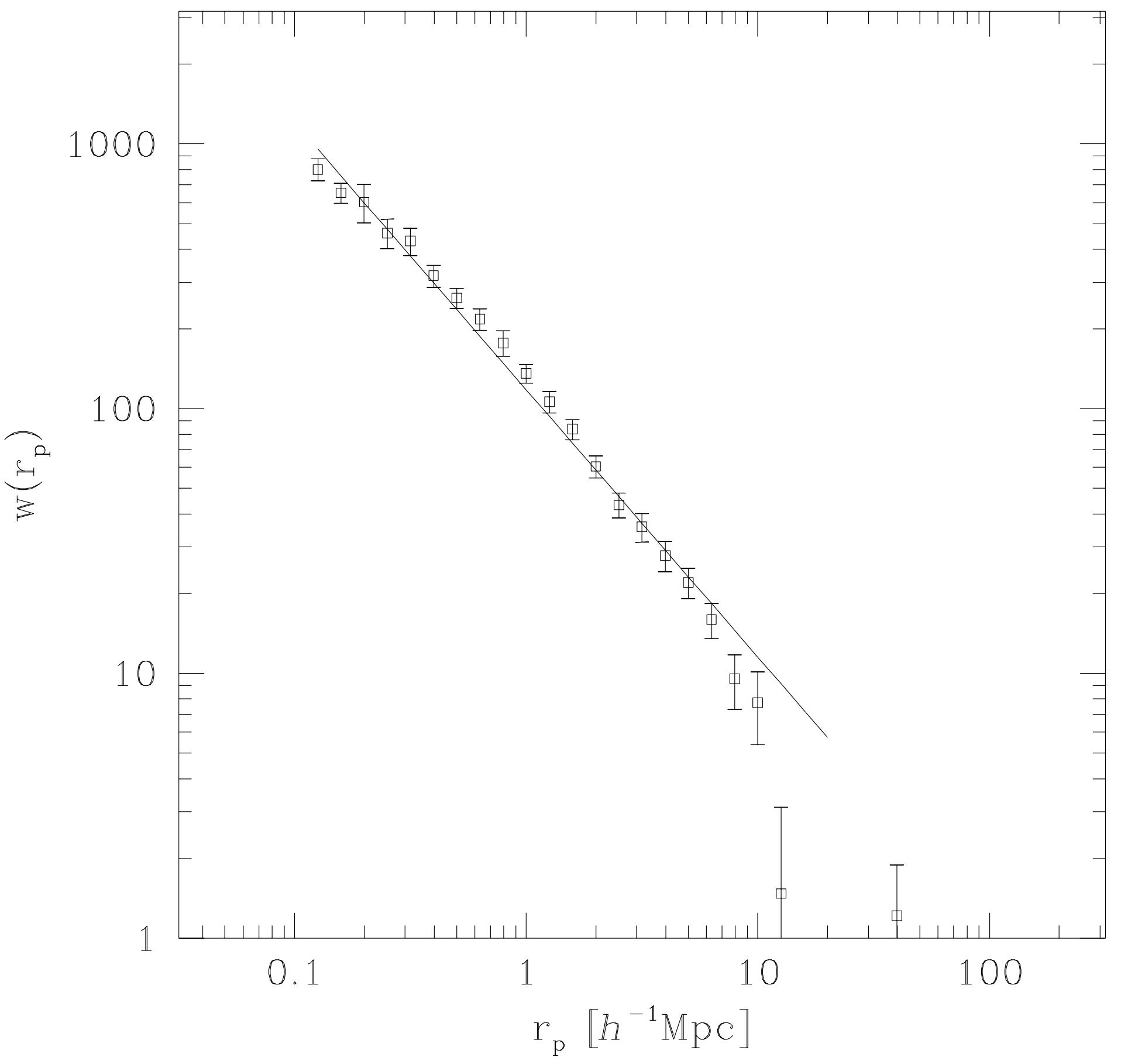

**Figure 4:** Projected correlation function $w(r_p)$ for the N-body simulation. The solid line is the power-law fit.

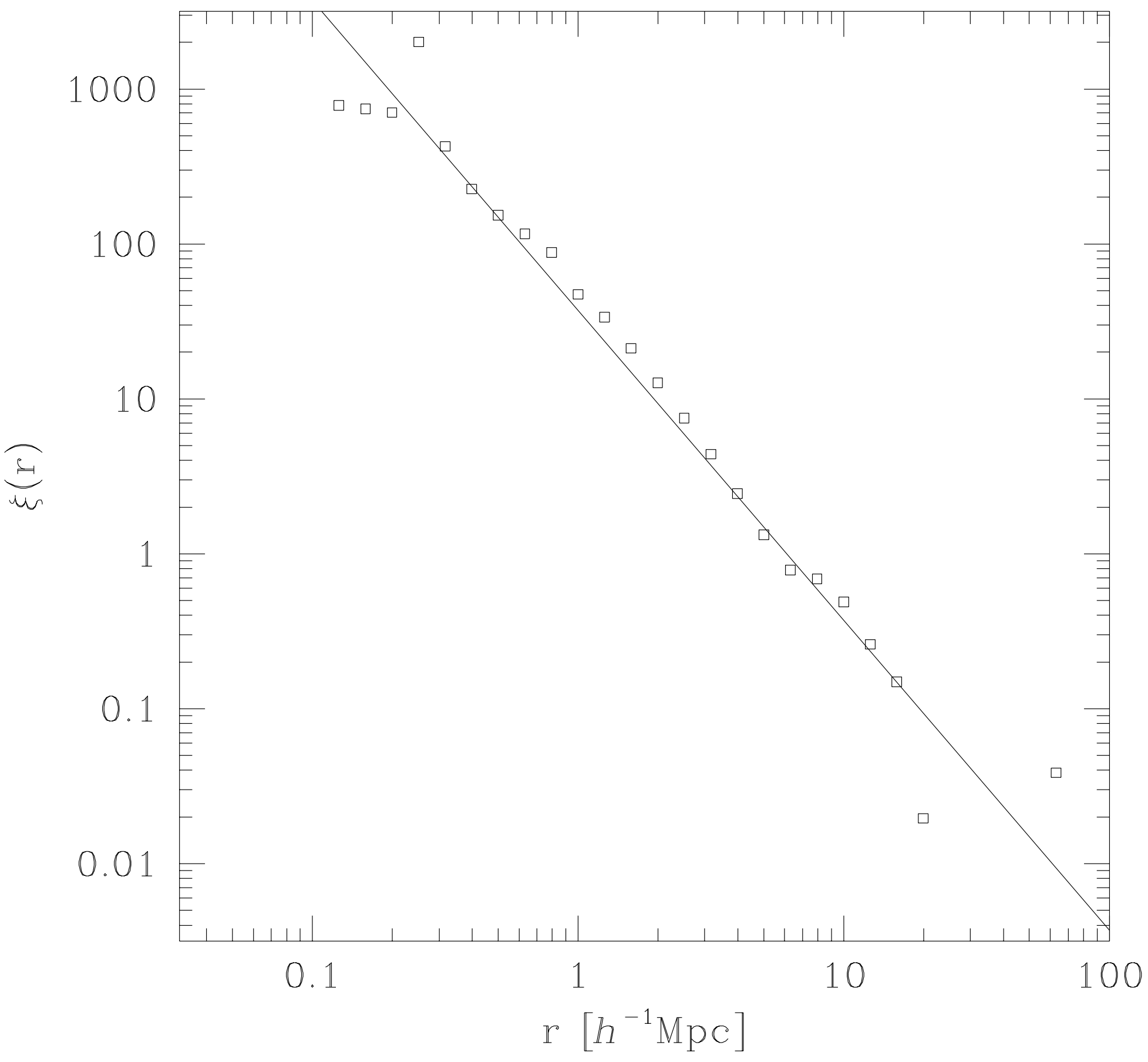

**Figure 5:** Two-point correlation function from pair counts in the N-body simulation. Solid line is the inversion of the power-law fit to $w(r_p)$.

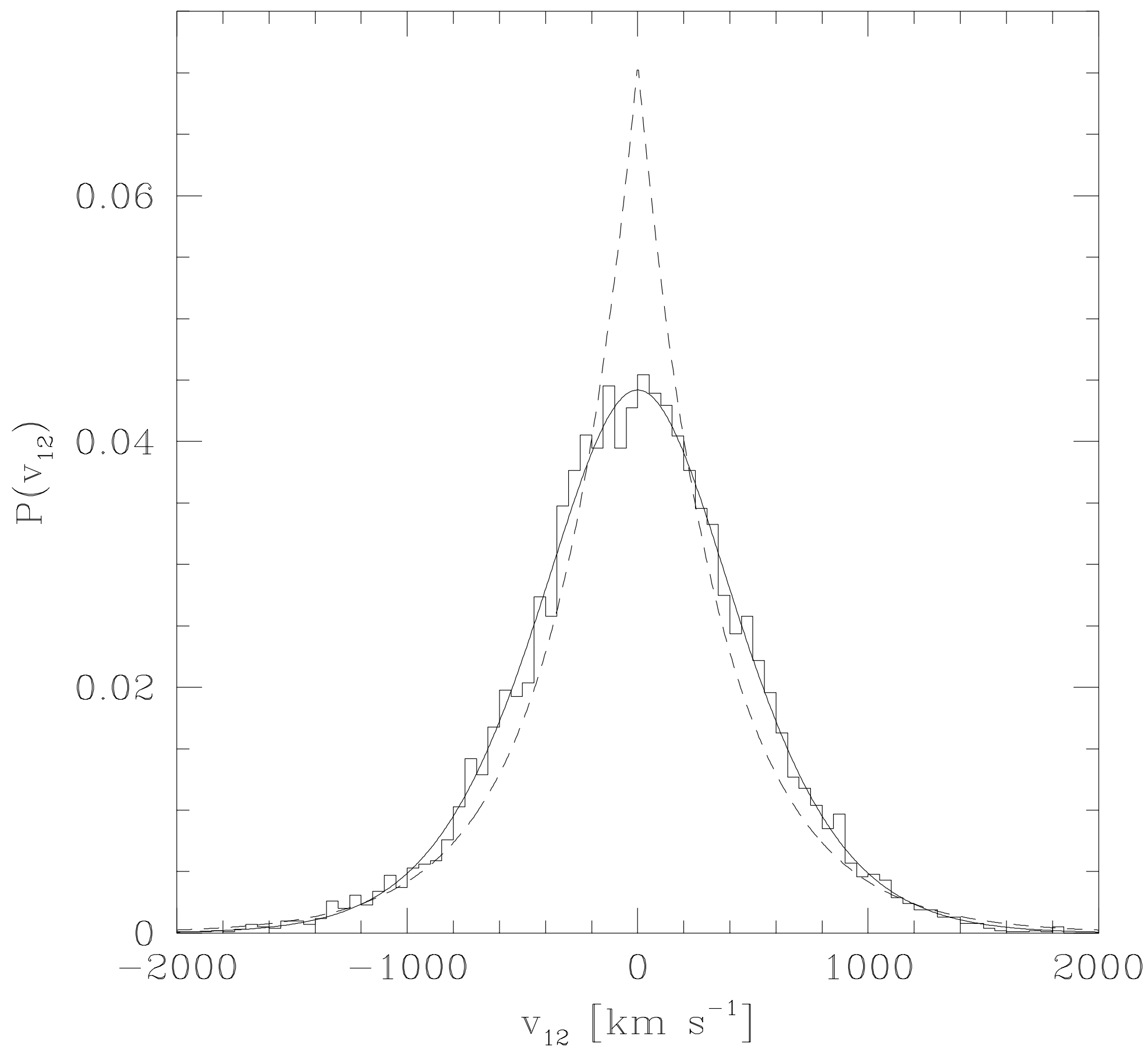

**Figure 6:** Distribution of pairwise velocities from Monte-Carlo simulations. Particle positions are from the N-body simulation. N-body velocities are replaced with velocities drawn randomly from the distribution of single-galaxy velocities in Equation 15. Magnitudes of the velocities are exponentially distributed with $\sigma_1 = 250\,\mathrm{km\,s^{-1}}$; directions are random. The solid line is the distribution of pairwise velocities predicted by Equation 18. The dashed line is the exponential distribution of 1-d pairwise velocities with the same second moment, $\sigma_{12} = 2\sigma_1$.

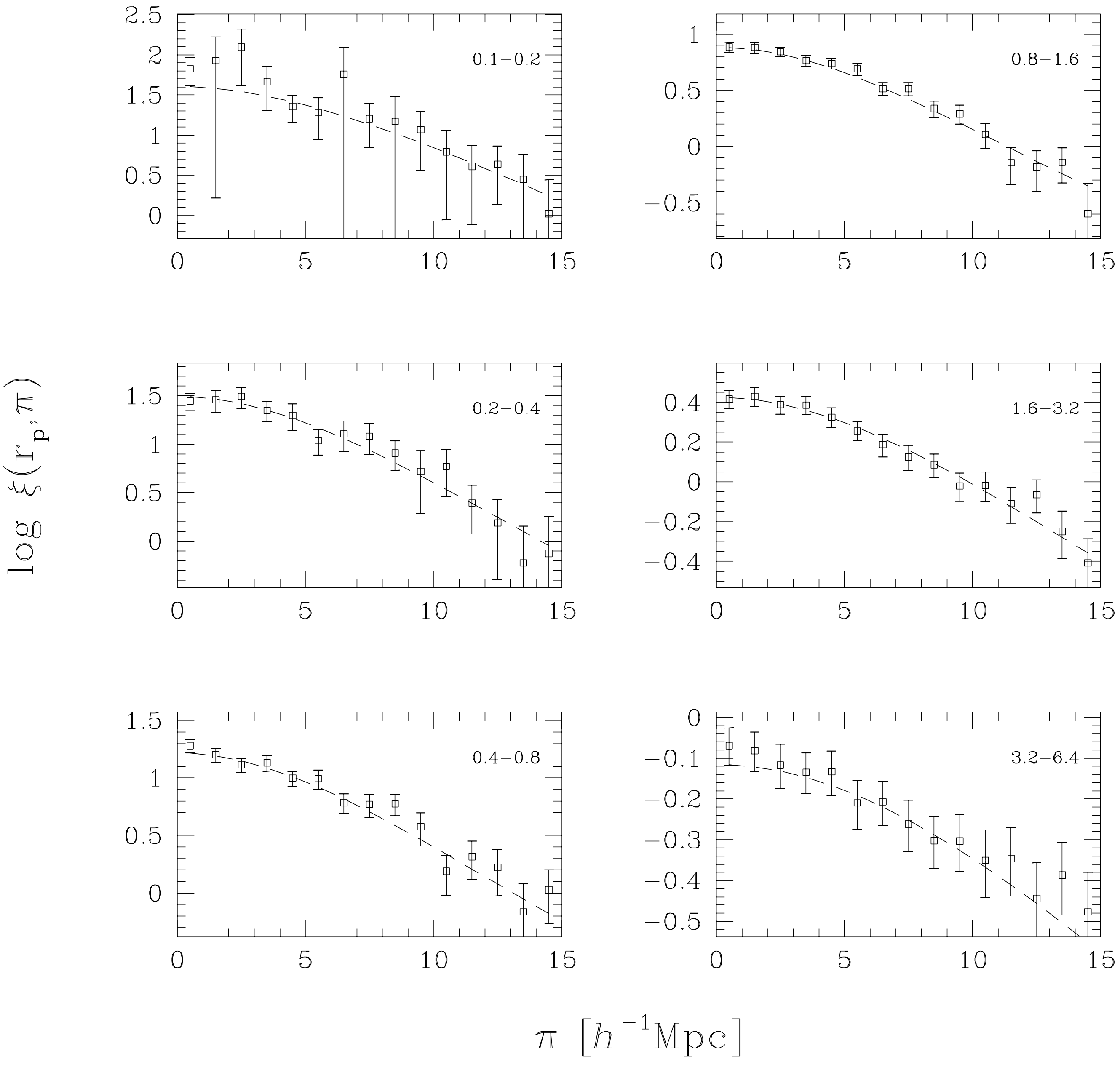

**Figure 7:** Fits to $\xi(r_p, \pi)$ derived from N-body simulation with velocities drawn from an isotropic exponential distribution (see figure 6). Numbers in the upper right-hand corner indicate the range of $r_p$ in $h^{-1}$ Mpc. Note that the vertical scale is different for each range of $r_p$.

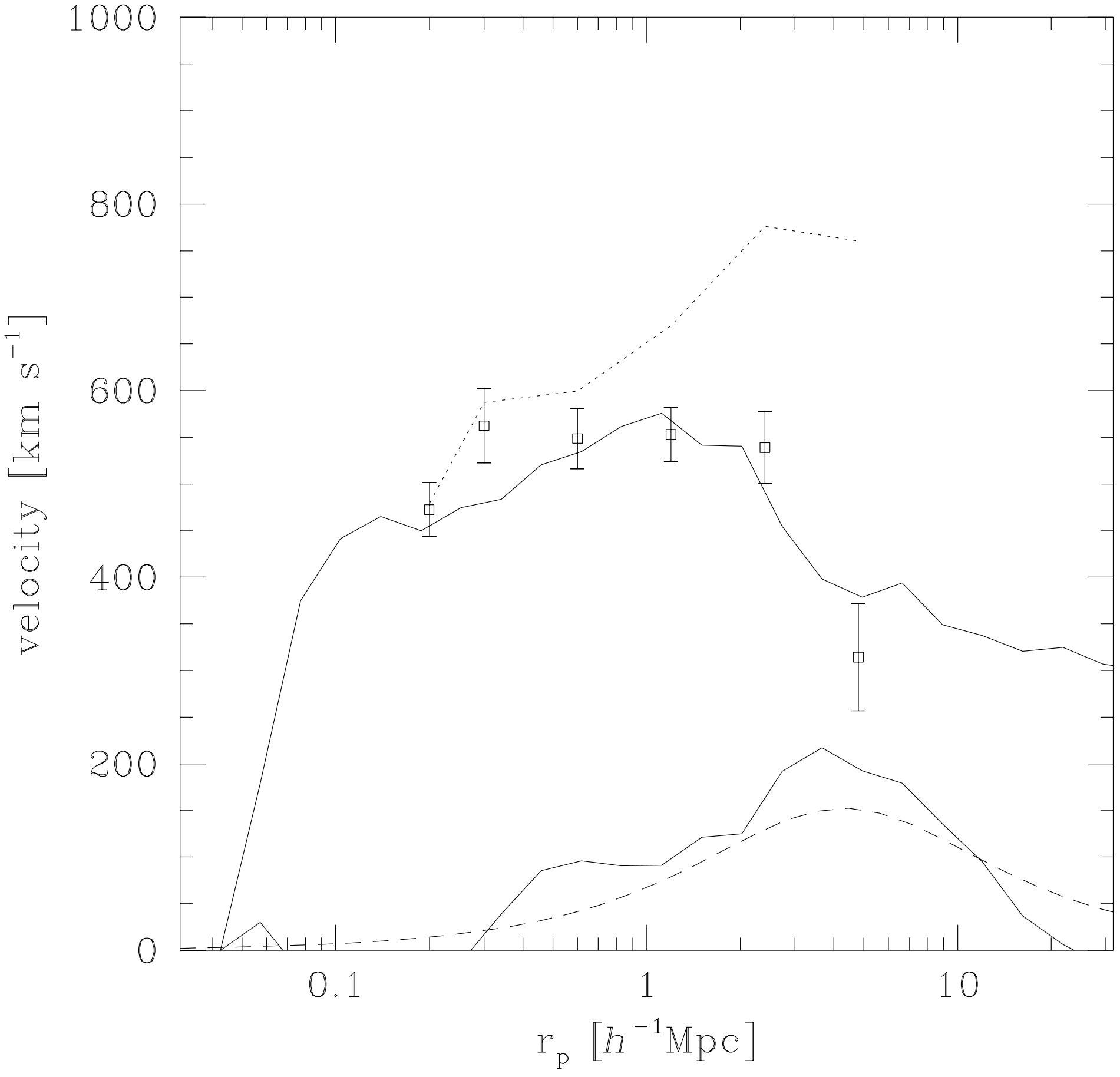

**Figure 8:** Solid lines are the first two central moments of the distribution $P(v_{12})$ for the N-body simulation. Open squares represent the fitted $\sigma_{12}$ ignoring infall. The dotted line is the measured $\sigma_{12}$ using the DP83 infall model (Equation 19) with $F = 1.5$, which is shown here as the dashed line.

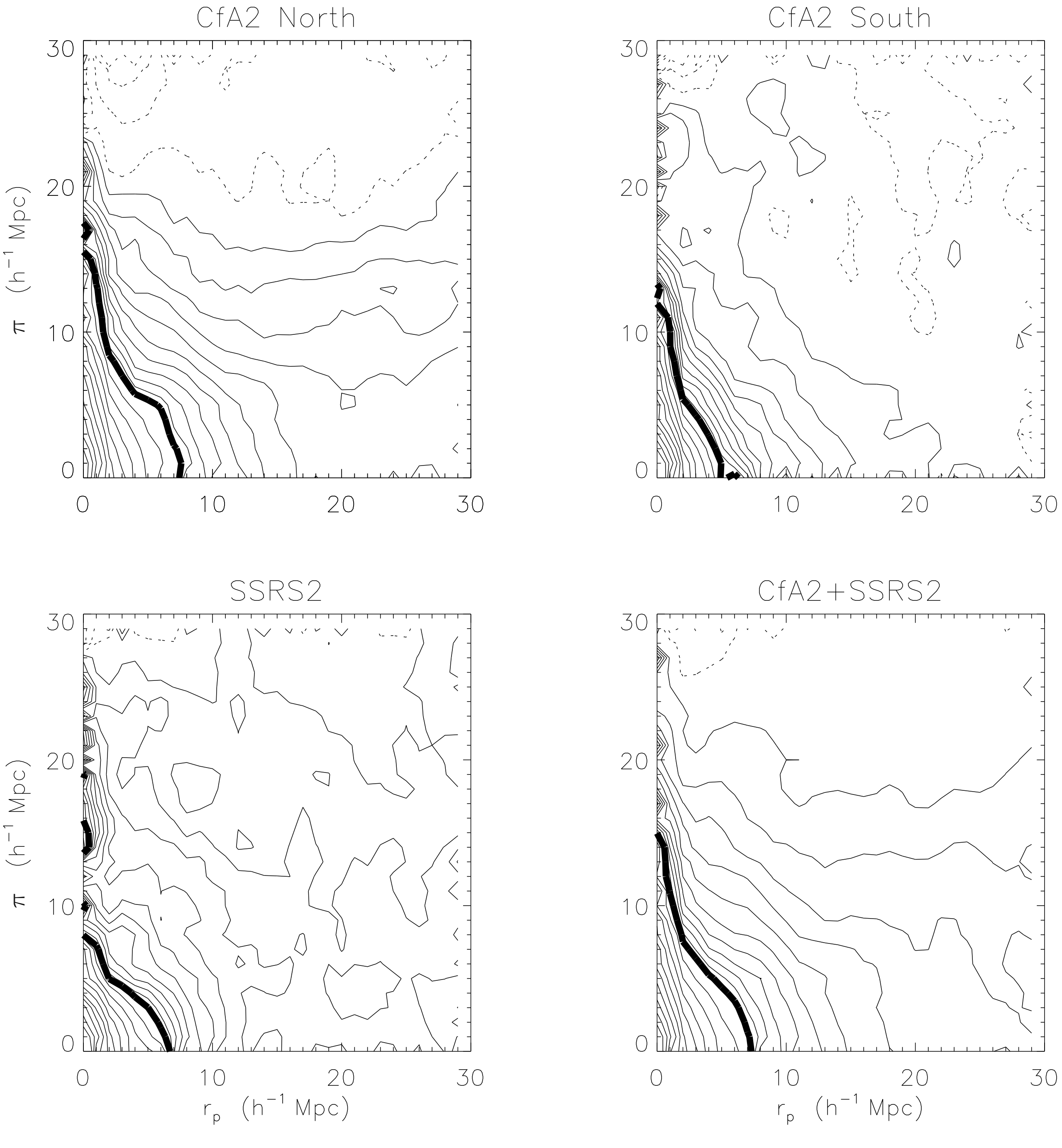

**Figure 9:** $\xi(r_p, \pi)$ for the four samples in Table 2 (MV weighting). The contour levels are described in the caption for Figure 2.

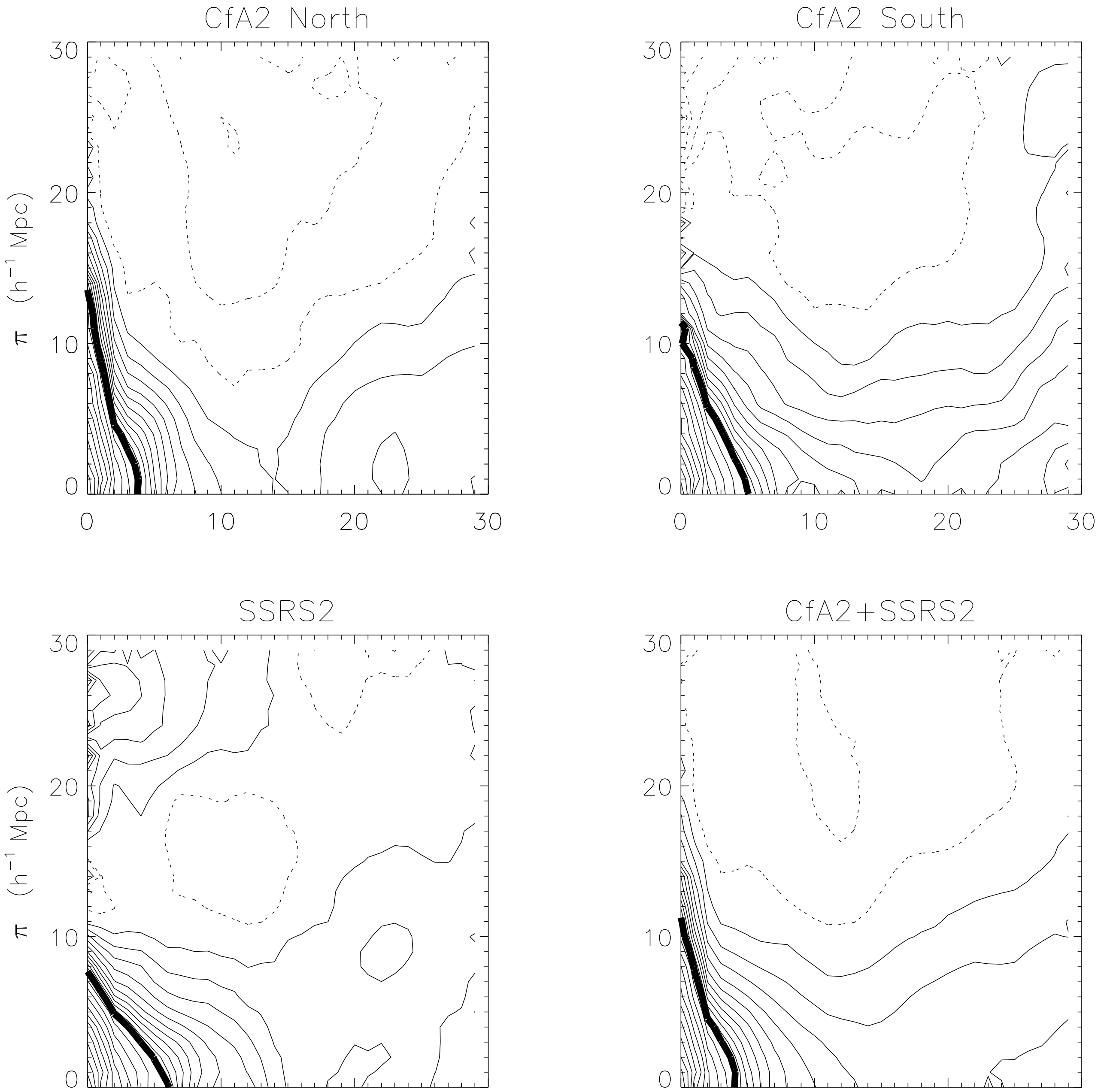

**Figure 10:** $\xi(r_p, \pi)$ for the four samples in Table 2 (uniform weighting). The contour levels are described in the caption for Figure 2.

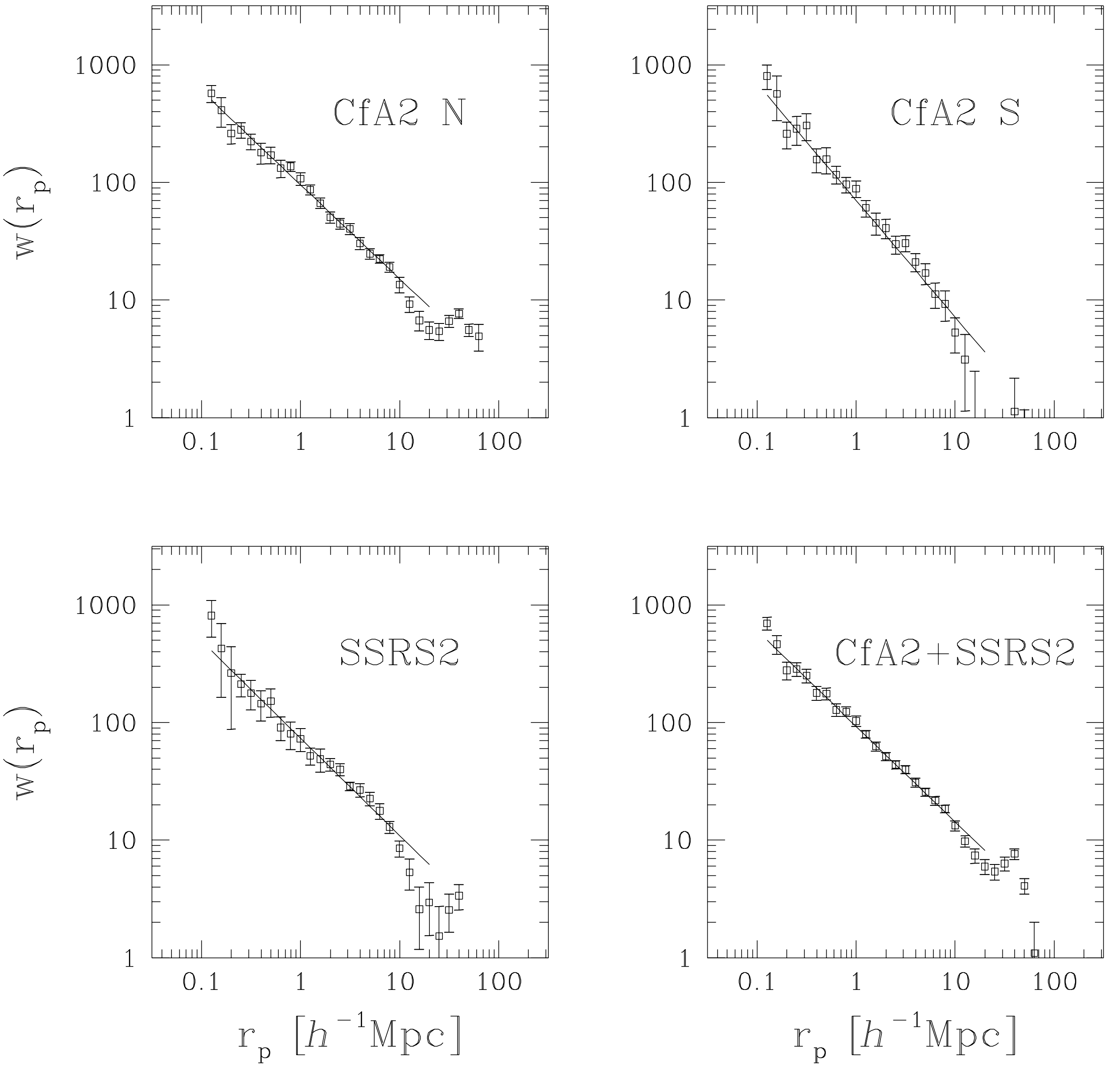

**Figure 11:** $w(r_p)$ for the four samples in Table 2. The solid line is the power-law fit given in Table 3.

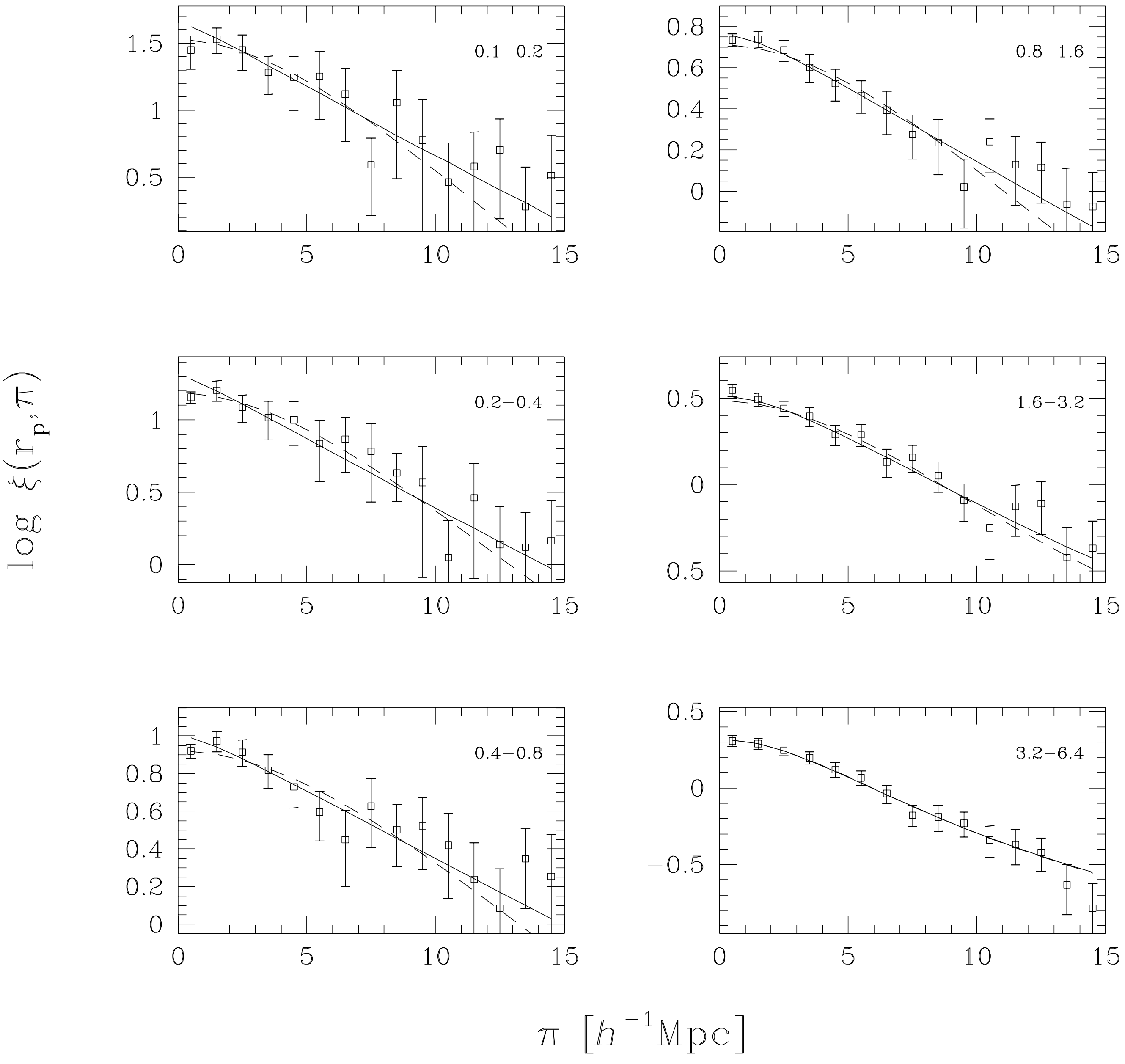

**Figure 12:** Fits to $\xi(\pi)$ at different $r_p$ for CfA2N. Solid line is the pairwise exponential; dashed line represents the single-galaxy exponential. Numbers in the upper right-hand corner indicate the range of $r_p$ in $h^{-1}$ Mpc. These fits do not include a correction for infall.

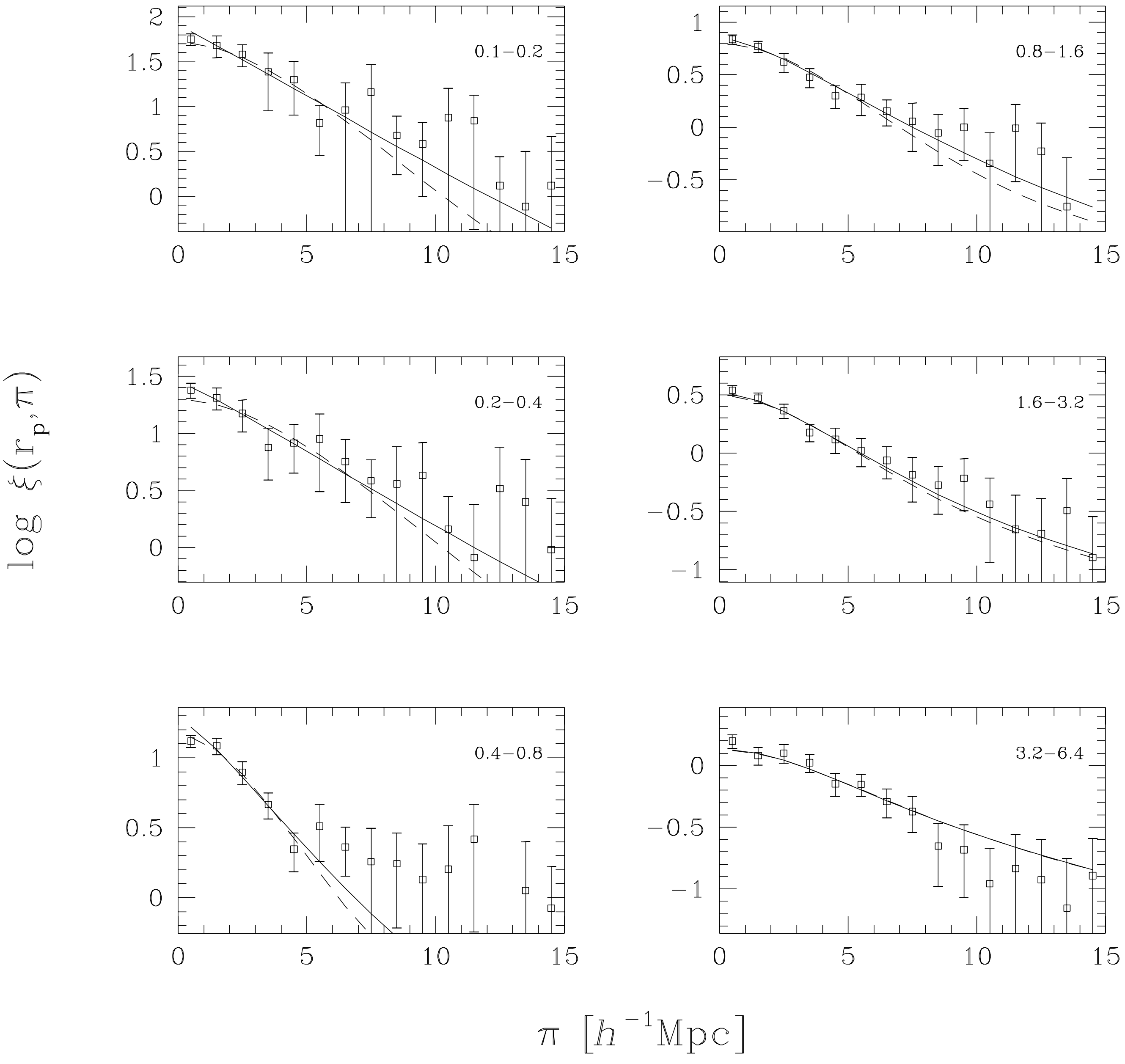

**Figure 13:** Fits to $\xi(\pi)$ at different $r_p$ for CfA2S. Solid line is the pairwise exponential; dashed line represents the single-galaxy exponential. Numbers in the upper right-hand corner indicate the range of $r_p$ in $h^{-1}$ Mpc. These fits do not include a correction for infall.

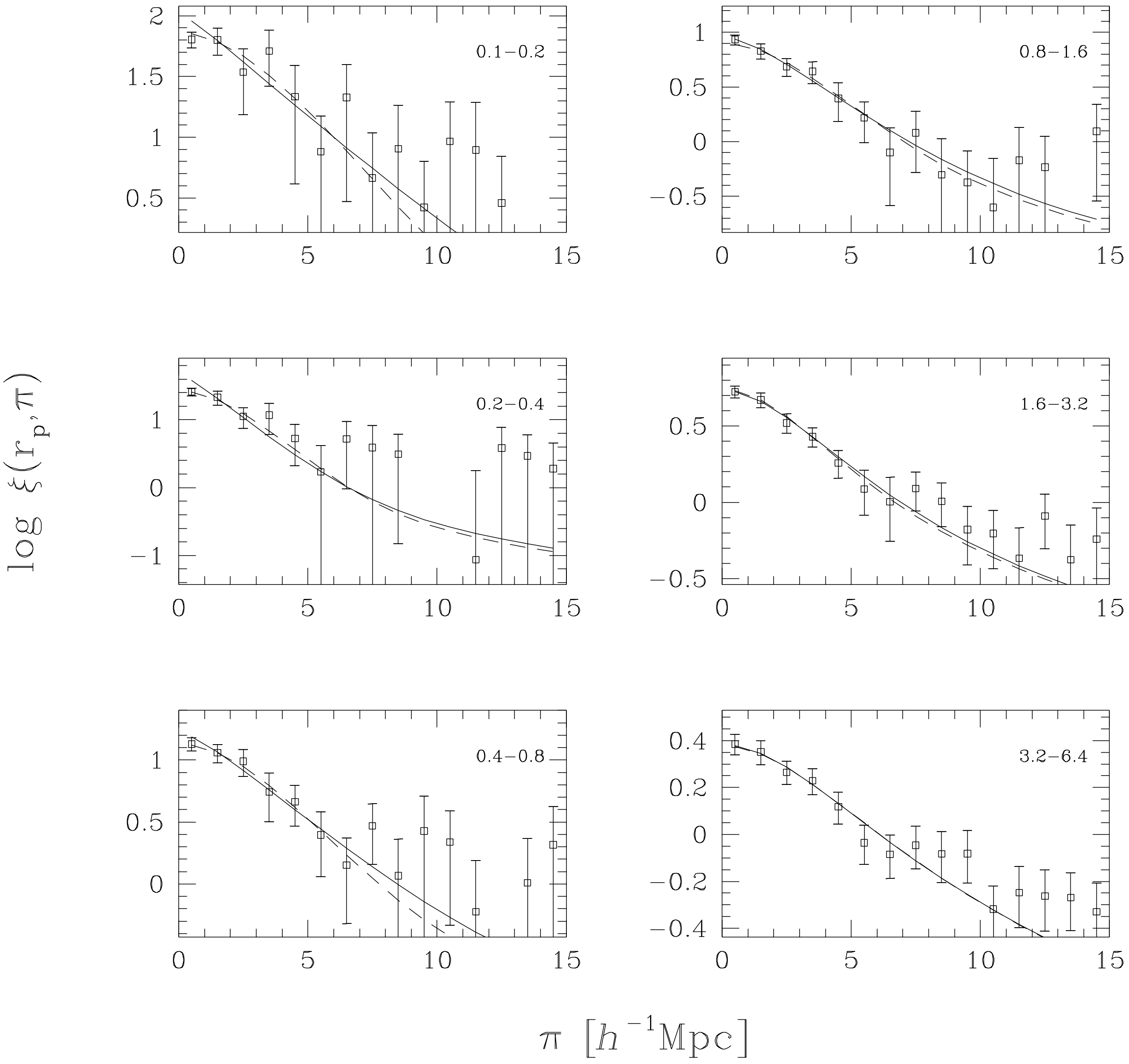

**Figure 14:** Fits to $\xi(\pi)$ at different $r_p$ for SSRS2. Solid line is the pairwise exponential; dashed line represents the single-galaxy exponential. Numbers in the upper right-hand corner indicate the range of $r_p$ in $h^{-1}$ Mpc. These fits do not include a correction for infall.

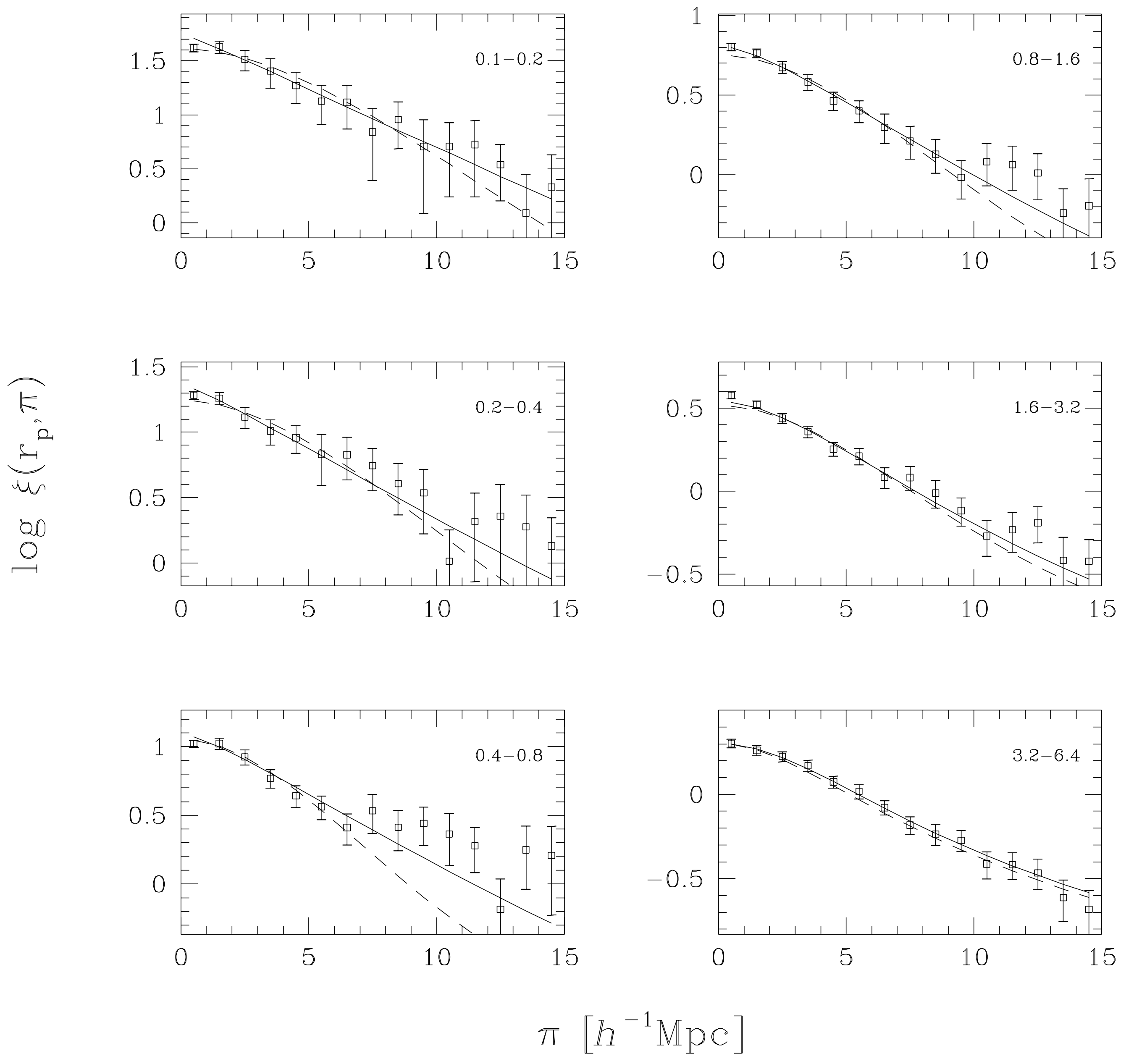

**Figure 15:** Fits to $\xi(\pi)$ at different $r_p$ for CfA2+SSRS2. Solid line is the pairwise exponential; dashed line represents the single-galaxy exponential. Numbers in the upper right-hand corner indicate the range of $r_p$ in $h^{-1}$ Mpc. These fits do not include a correction for infall.

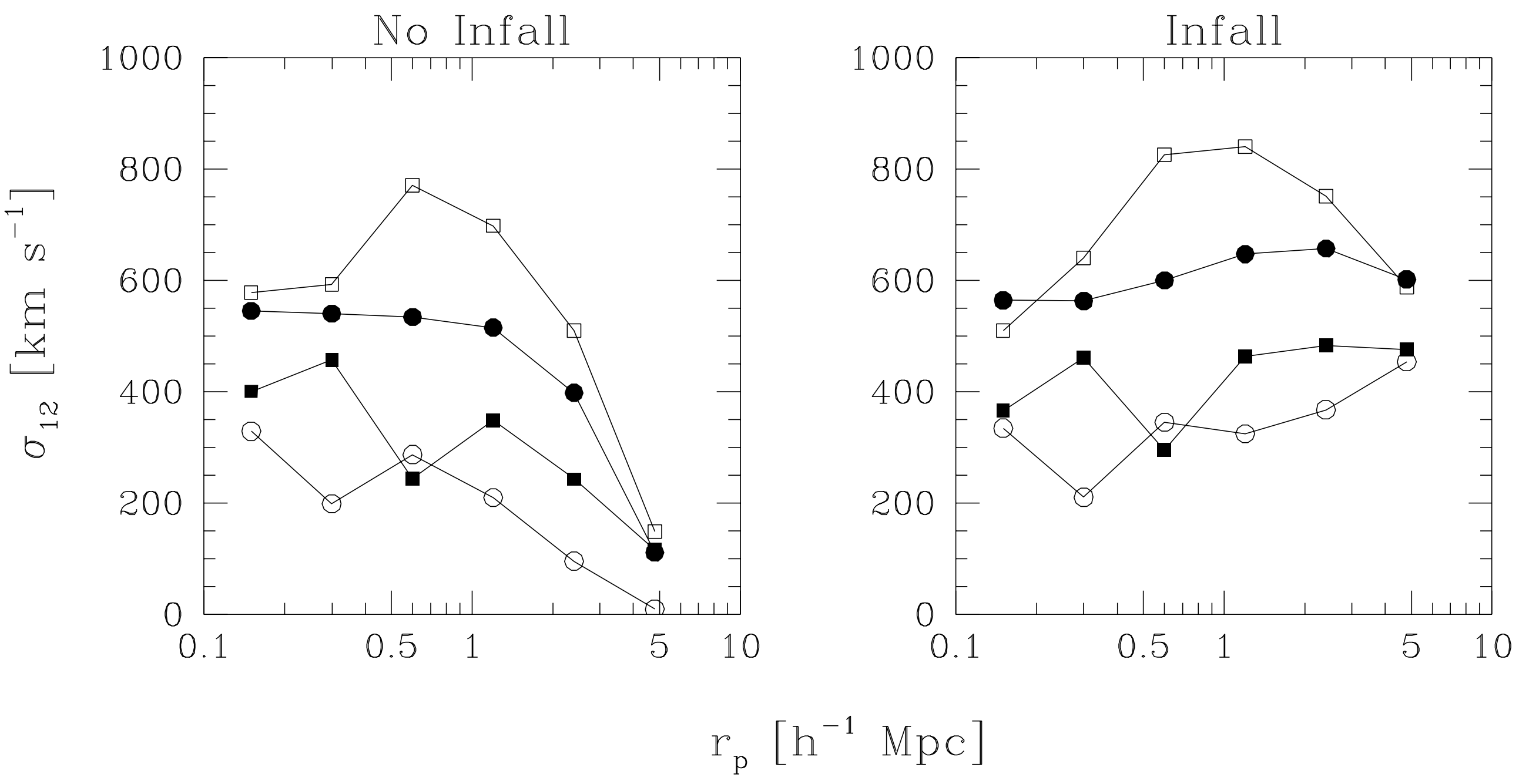

**Figure 16:** Scaling of $\sigma_{12}$ with $r_p$. The samples are CfA2N (open squares), CfA2S (filled squares), SSRS2 (open circles) and CfA2+SSRS2 (filled circles).

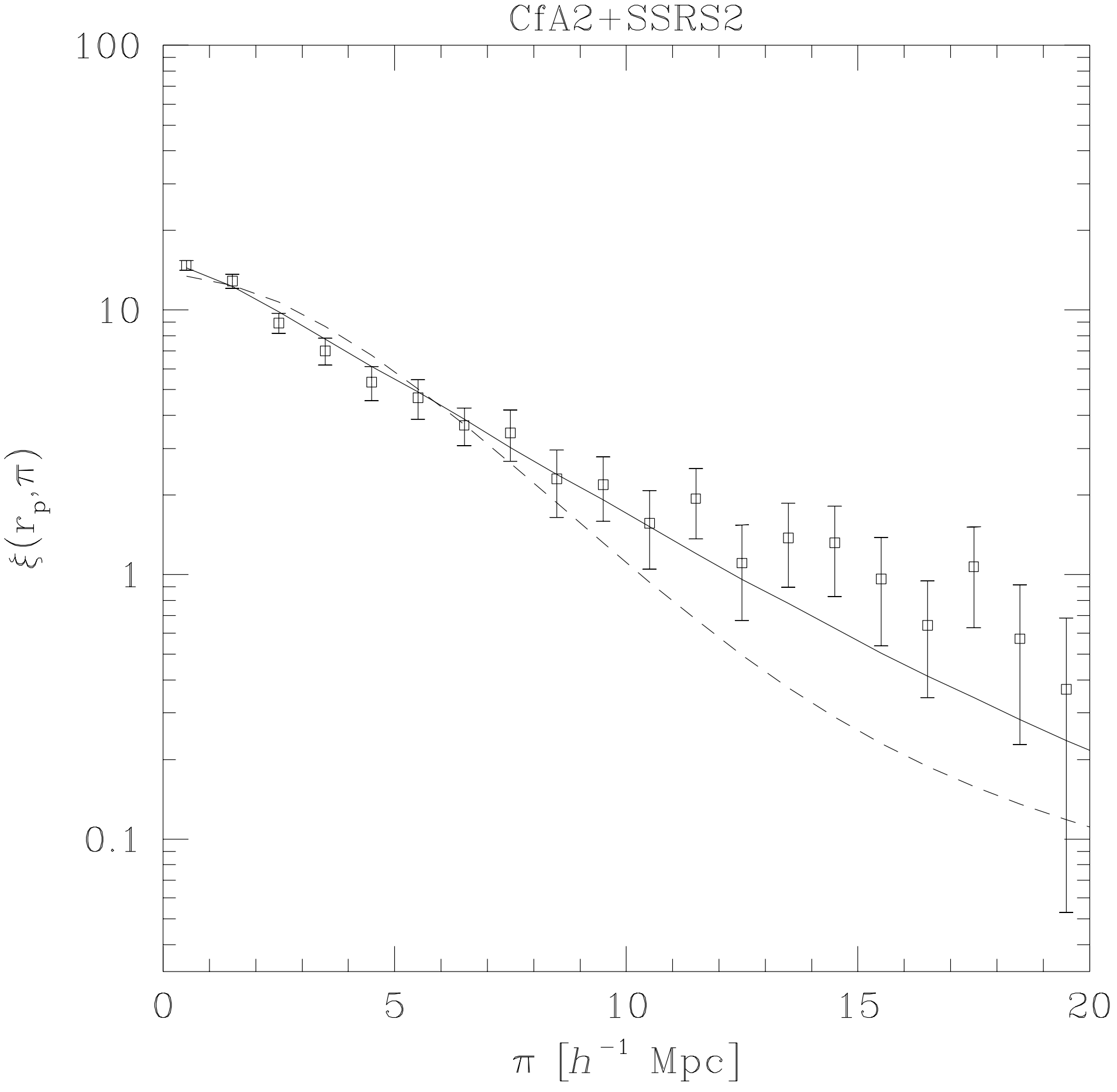

**Figure 17:** Fits of Equations 15 and 18 to $\xi(r_p, \pi)$ for $r_p \leq 1\,h^{-1}\,\mathrm{Mpc}$ in CfA2+SSRS2. Solid line is the pairwise exponential; dashed line represents the single-galaxy exponential.

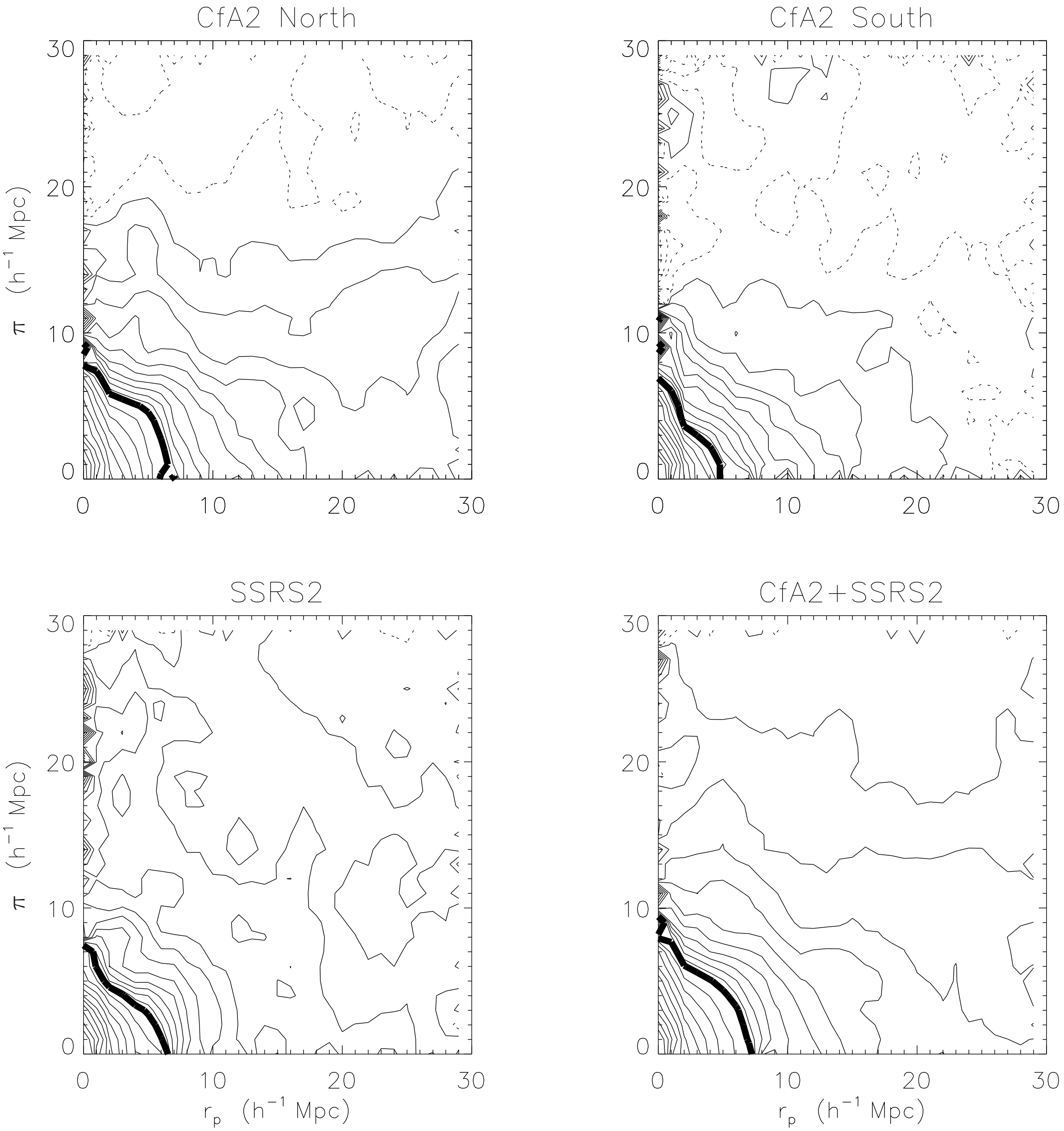

**Figure 18:** $\xi(r_p, \pi)$ for the four samples in Table 2 after removing Abell clusters with $R \geq 1$. The contour levels are described in the caption for Figure 2.

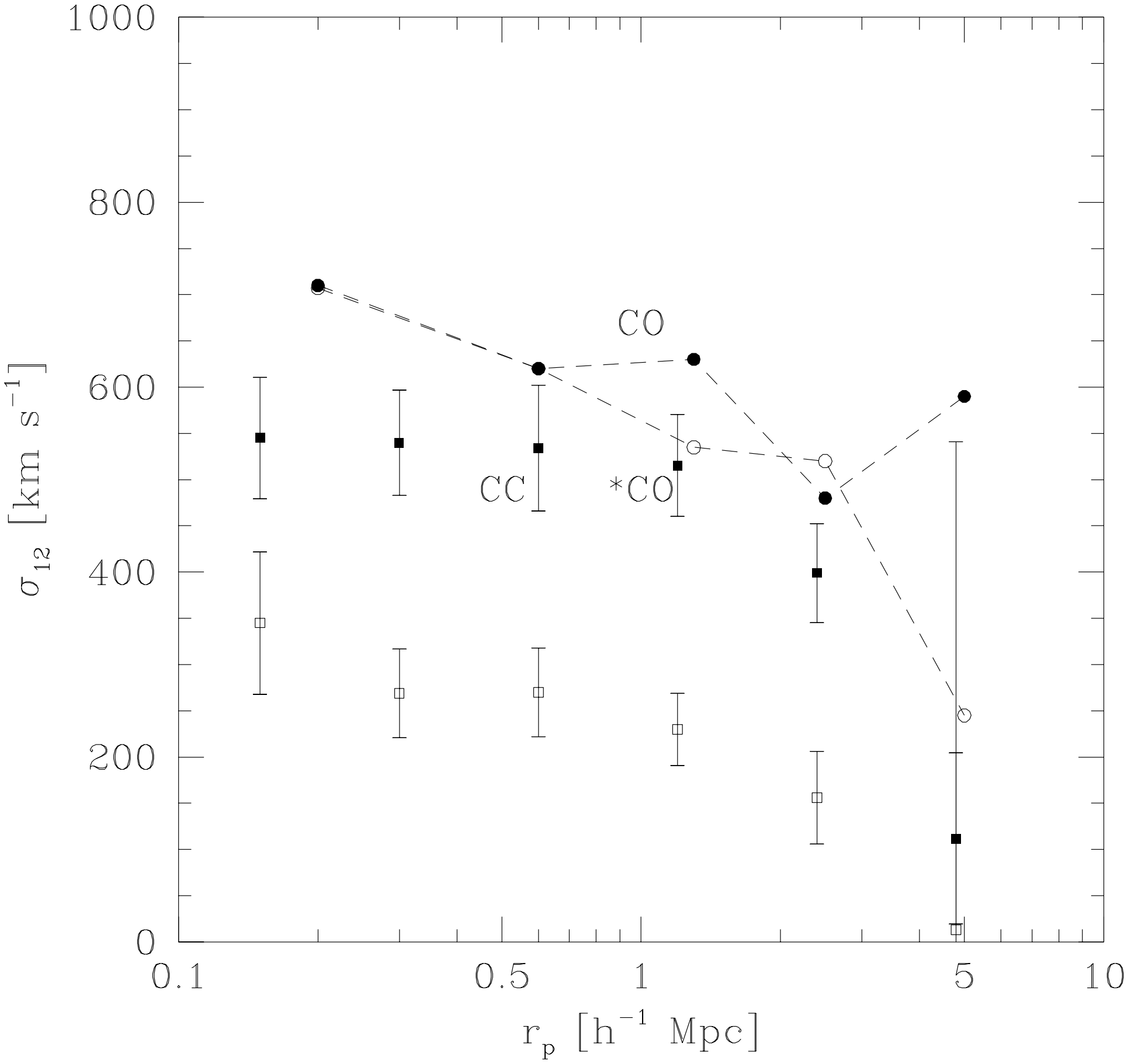

**Figure 19:** Comparison with $\sigma_{12}$ derived from N-body simulations. The filled squares are our results for CfA2+SSRS2 with no correction for infall. Open squares represent CfA2+SSRS2 with Abell clusters removed. Filled circles are from Davis *et al.* (1985), where we have taken the mean of the two simulations from their figure 15. CC comes from Couchman & Carlberg (1992). CO represents Cen & Ostriker (1993); $^*$CO represents their "bright" sample, which includes only galaxies with mass greater than $1.1 \times 10^{10} M_\odot$.

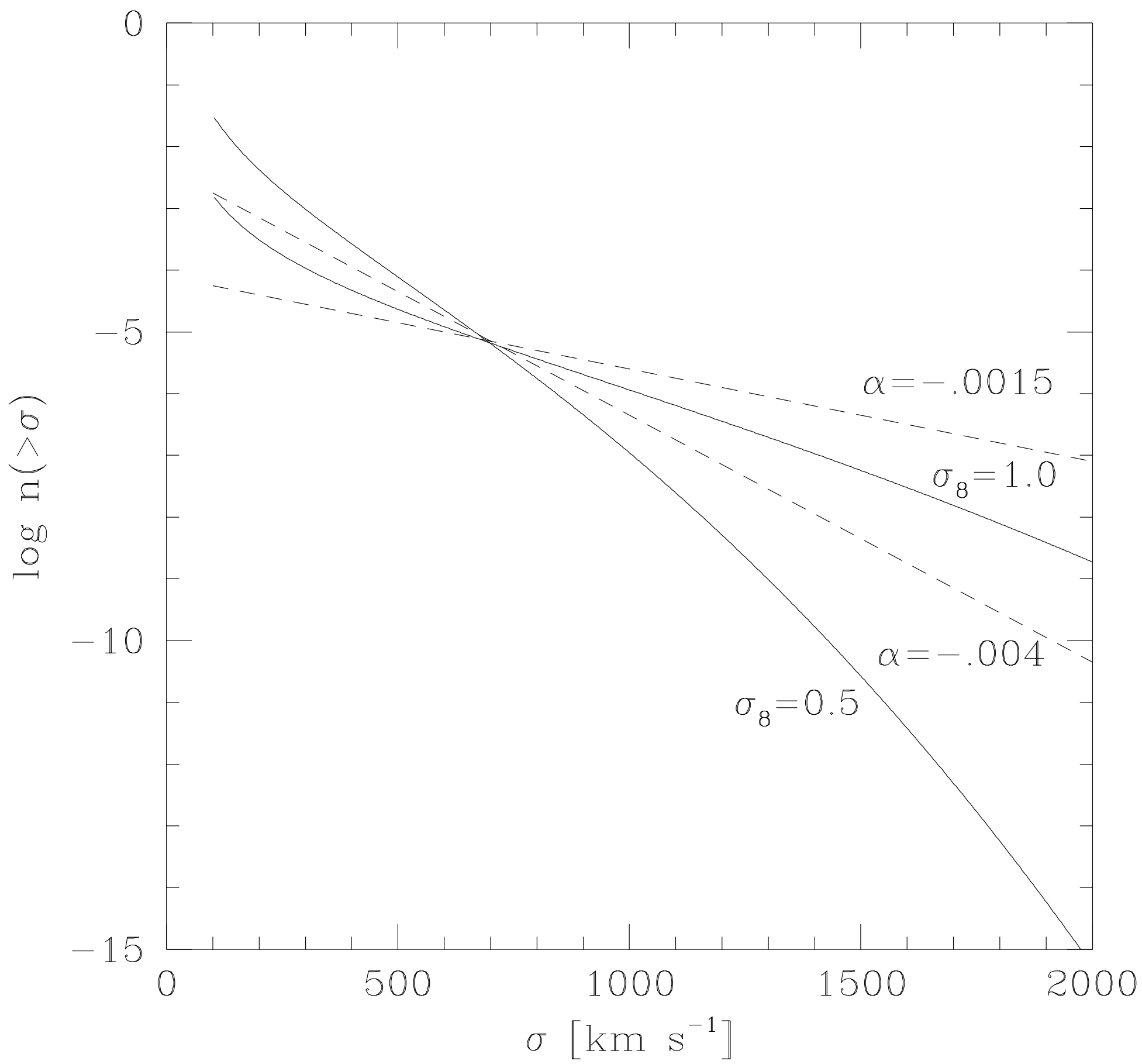

**Figure 20:** Cumulative distribution of one-dimensional velocity dispersions $n(>\sigma)$ for systems of galaxies. The curves labeled by $\sigma_8$ are the predictions of the Press-Schechter prescription; $\sigma_8$ is the normalization of a flat CDM power spectrum. Curves labeled by $\alpha$ represent the range of distributions $n(>\sigma) \propto 10^{\alpha\sigma}$ derived from optical surveys. X-ray observations correspond best with $\sigma_8 = 0.6$ for flat CDM. All distributions are normalized to match the observed number density of Abell clusters with richness class $R \geq 1$.

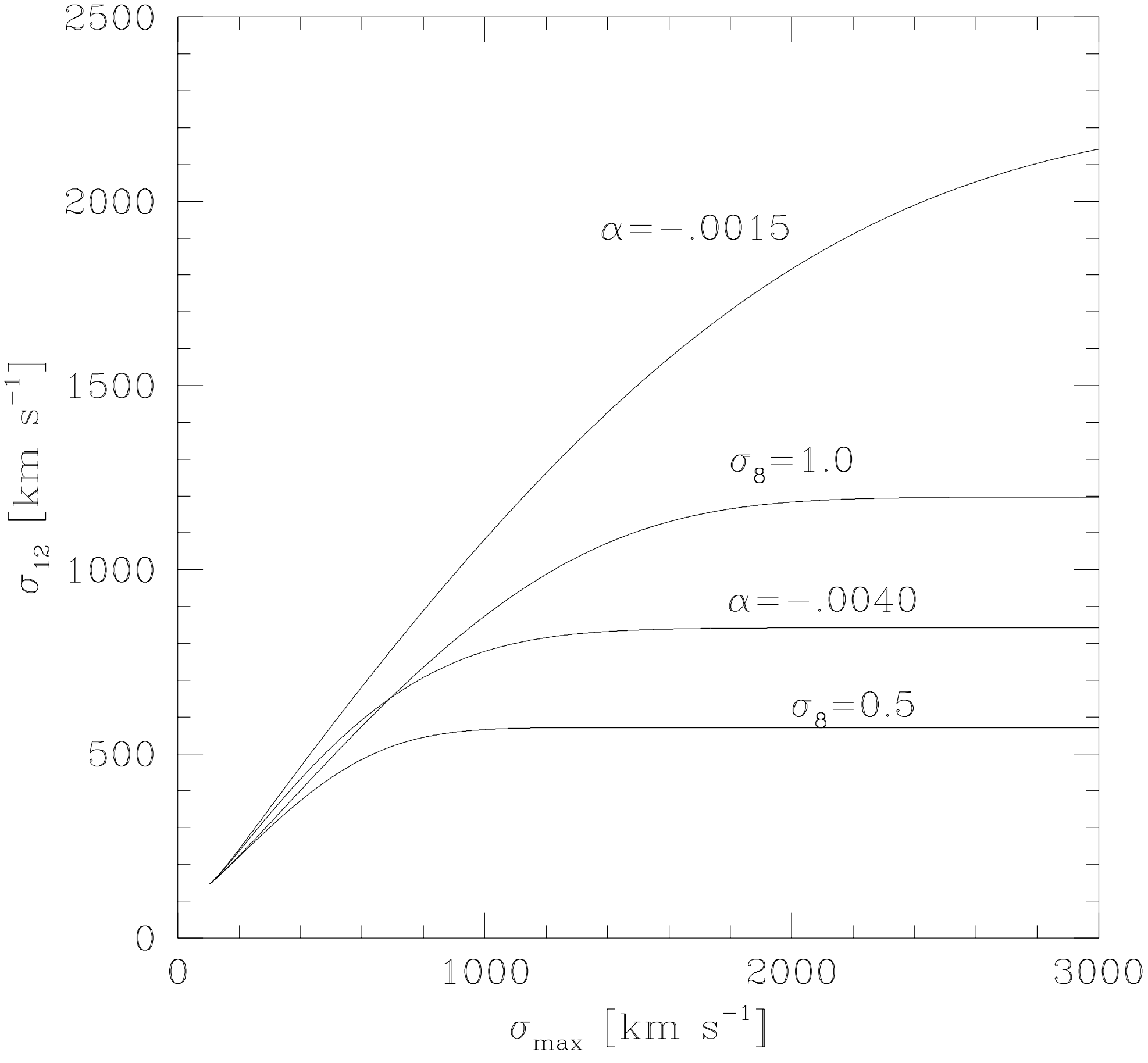

**Figure 21:** Convergence of $\sigma_{12}$ averaged over virialized systems. $\sigma_{max}$ is the upper limit of integration in Equation 22. Curves correspond to the distributions $n(>\sigma)$ in figure 20 and are labeled accordingly.

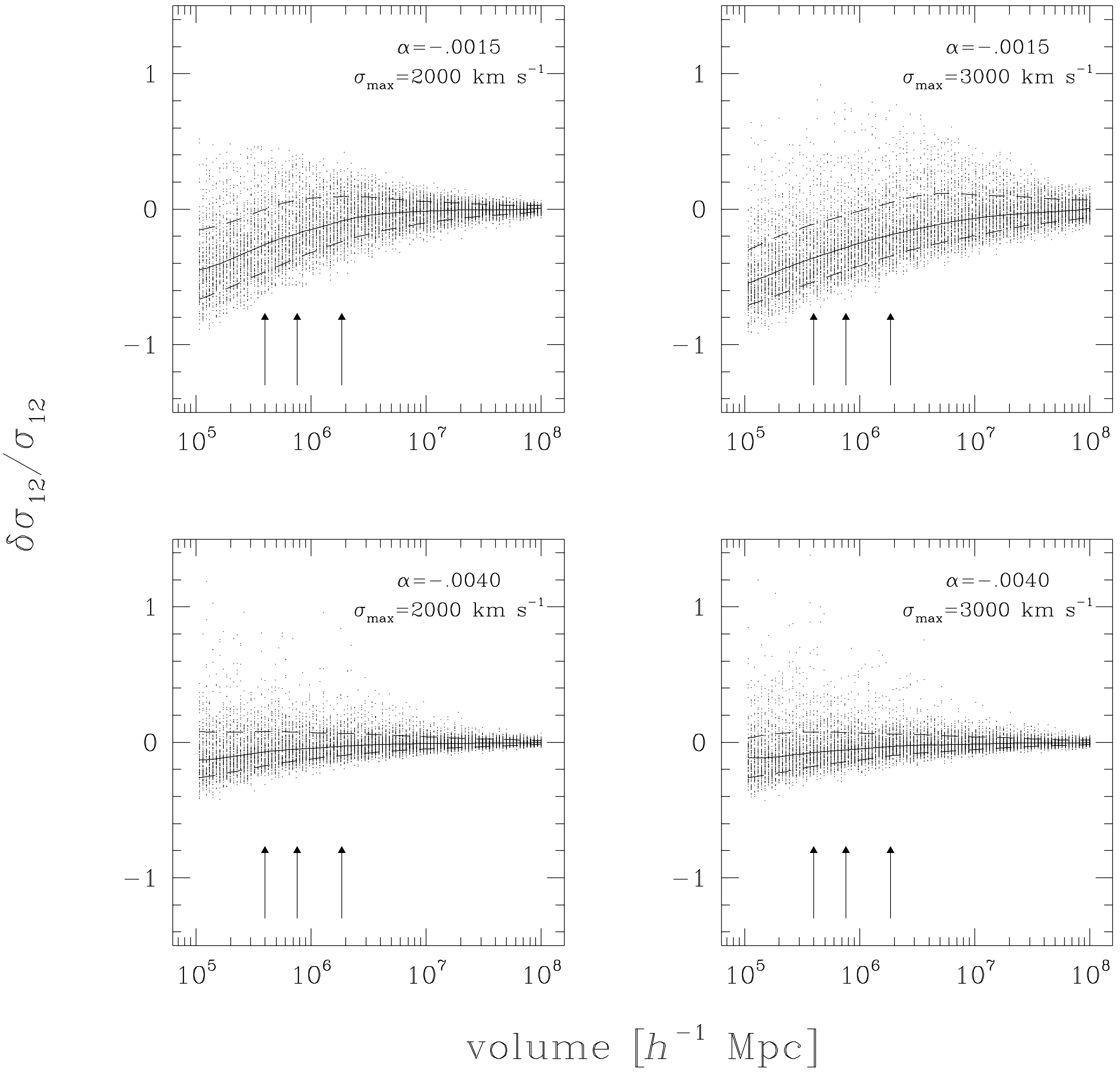

**Figure 22:** Monte-Carlo analysis of the error in $\sigma_{12}$ as a function of survey volume. The pairwise velocity dispersion is biased low when the volume does not include a significant number of clusters with $\sigma > \sigma_{conv}$, where $\sigma_{conv}$ is the velocity dispersion $\sigma$ at which $\sigma_{12}$ converges (see figure 21). The dispersion $\sigma_{max}$ is the upper limit of integration in Equation 22. The slopes $\alpha$ of the exponential $n(\sigma)$ correspond to the observed distributions in figure 20. Arrows indicate (from left to right) the volumes of CfA1, CfA2N and CfA2+SSRS2.